%% file: main.tex
\documentclass[a4paper,fleqn]{cas-dc}

\usepackage[numbers]{natbib}
\usepackage{enumitem}
\usepackage{pifont}
\usepackage{placeins}
\usepackage{tabularx}

\graphicspath{{./figs/}}

\begin{document}
\let\WriteBookmarks\relax
\def\floatpagepagefraction{1}
\def\textpagefraction{.001}

\shorttitle{A Systematic Survey of Decompilation}
\shortauthors{Shimane et al.}

\title[mode=title]{A Systematic Survey of Decompilation: Mapping Classical and Machine-Learning Approaches}

\author[1]{Keitaro Shimane}
\author[1]{Minami Yoda}
\cormark[1]
\ead{yoda.minami@nihon-u.ac.jp}
\author[2]{Jialong Li}
\author[1]{Yutaka Matsuno}

\cortext[1]{Corresponding author.}
% Contact information can be added once the corresponding author is fixed.
% Example:

\affiliation[1]{
  organization={Nihon University},
  city={1-21-1 Nishi-Waseda, Shinjuku-ku, Tokyo},
  country={Japan}}

\affiliation[2]{
  organization={Waseda University},
  city={7-24-1 Narashinodai, Funabashi, Chiba},
  country={Japan}}

% Detailed affiliation candidates, kept for later confirmation.
% \author[1]{Keitaro Shimane}
% \author[2]{Minami Yoda}
% \author[3]{Jialong Li}
% \author[2,1]{Yutaka Matsuno}
%
% \affiliation[1]{
%   organization={Department of Computer Science, Graduate School of Science and Technology, Nihon University},
%   city={Chiba},
%   country={Japan}}
%
% \affiliation[2]{
%   organization={Department of Computer Engineering, College of Science and Technology, Nihon University},
%   city={Chiba},
%   country={Japan}}
%
% \affiliation[3]{
%   organization={Waseda Institute for Advanced Study, Waseda University},
%   city={Tokyo},
%   country={Japan}}

\begin{abstract}
\noindent\textbf{Context:}
Decompilers reconstruct source-level code from compiled binaries, and
analysts depend on them whenever source code is unavailable. Many recent
studies apply machine learning (ML), including large language models, to
parts of this task. From the primary literature alone, it is hard to tell
where these methods sit in the classical decompilation pipeline and how
well their claims are supported.

\noindent\textbf{Objectives:}
We map decompilation research onto a pipeline-centered taxonomy and ask
two questions: where classical, ML-trained, and ML-prompted approaches
are positioned (RQ1), and how ML-based studies vary in technical design
and evaluation evidence (RQ2).

\noindent\textbf{Methods:}
We systematically surveyed 405 decompilation studies published between
1973 and 2026, collected through database search and citation tracing.
We classified 394 primary studies into eight pipeline stages and two
complementary categories. For the 116 ML-based studies that propose or
modify decompilation methods, we further extracted the model-use
strategy, reusable backbones, target platform, and the evidence behind
each evaluation.

\noindent\textbf{Results:}
Classical techniques still dominate front-end recovery, IR lifting, and
control-flow structuring (277 of 394 primary studies are classical).
ML-based work (117 studies) clusters instead around symbol and name
recovery, output refinement, and end-to-end translation. Type recovery
is the clearest mixed region. Most ML-based studies target native
binaries (97 of 116), and most support their claims with item-level
matching against references (89 of 116); execution-based tests, human or
LLM judgment, and downstream-task evidence are much rarer.

\noindent\textbf{Conclusion:}
ML has not displaced the classical pipeline. It has been taken up where
a recovered artifact can be posed as a prediction, translation, or
refinement problem, and reported gains only make sense when read against
the pipeline position and the evidence used to support them.
\end{abstract}

%\begin{highlights}
%\item Maps 405 decompilation studies across a pipeline-centered taxonomy.
%\item Compares classical, ML-trained, and ML-prompted decompilation work.
%\item Reviews ML-based design patterns and evaluation evidence.
%\item Identifies where ML enters the decompilation pipeline.
%\end{highlights}


\maketitle

\input{sections/1_introduction}
\input{sections/2_background}
\input{sections/3_methodology}
\input{sections/5_landscape}
\input{sections/6_focused_review}
\input{sections/7_discussion}
\input{sections/8_threats}

\input{sections/9_conclusion}

\section*{CRediT authorship contribution statement}
\textbf{Keitaro Shimane:} Conceptualization, Methodology, Investigation,
Data curation, Writing -- original draft.
\textbf{Minami Yoda:} Validation, Writing -- review \& editing.
\textbf{Jialong Li:} Writing -- review \& editing.
\textbf{Yutaka Matsuno:} Supervision, Writing -- review \& editing.

\section*{Declaration of competing interest}
The authors declare that they have no known competing financial
interests or personal relationships that could have appeared to
influence the work reported in this paper.

\section*{Data availability}
The classification and extraction records for all included studies are
available in the replication package at
[URL to be inserted upon acceptance]. % TODO: replace with archived DOI/URL before submission.

\bibliographystyle{cas-model2-names}
\bibliography{references}

\end{document}

%% file: sections/1_introduction.tex
\section{Introduction}\label{sec:introduction}

Decompilation, the recovery of high-level source code from compiled
binaries, is a cornerstone capability for security analysis, malware
reverse engineering, vulnerability discovery, and legacy software
comprehension.
When source code is unavailable, analysts depend on decompilers to
transform opaque machine code into a form that can be read, reasoned
about, and acted upon.
The practical importance of decompilation is reflected in the
widespread adoption of tools such as IDA~Pro/Hex-Rays,
Ghidra, and Binary~Ninja across both industry and academia.

The field has a long history.
The earliest decompilation efforts appeared in the 1960s as tools for
cross-architecture program migration~\cite{CifuentesPhD1994,Housel1974}.
Cifuentes's doctoral thesis in 1994~\cite{CifuentesPhD1994} provided
a widely used systematic framework for native-binary decompilation,
decomposing the process into a machine-dependent front-end, a universal
analysis core, and a target-language back-end.
The release of the Hex-Rays Decompiler in 2007 established C-like
pseudocode as the output standard for reverse engineering practice,
and the open-sourcing of Ghidra in 2019 gave researchers broad access
to production-quality decompiler infrastructure.
Across this history, classical techniques grounded in compiler theory,
static analysis, and formal methods have achieved reliable recovery
of control flow, intermediate representations, and basic type
information.
While these methods provide structurally sound recovery, they often
struggle to reconstruct the semantic intent---such as original variable
names and high-level abstractions---that is lost during compilation.

To address this semantic gap, machine-learning-based approaches have
increasingly entered the field.
Task-trained neural models have demonstrated strong results in
variable naming~\cite{Dramko2023}, type inference~\cite{DIRTY2022},
and symbol recovery~\cite{SymLM2022}, while large language models
have been applied to end-to-end code
generation~\cite{LLM4Decompile2024} and output
refinement~\cite{DecLLM2025}.
Yet the impact of ML on decompilation has been uneven:
certain pipeline stages (e.g., variable naming and
output refinement) have seen rapid ML adoption, while
others (e.g., control-flow structuring and IR lifting) remain
predominantly classical.
The question that arises is where in the decompilation pipeline
ML has been applied, where it has not, and why.

Seeing the field as a whole is difficult.
The literature spans security, programming languages, software
engineering, and artificial intelligence venues, so no single
disciplinary perspective covers it well.
The pipeline itself has evolved unevenly: some stages are mature and
well-studied, whereas others have only recently emerged as independent
research areas.
At the same time, ML-based work, especially work involving large
language models, is moving quickly enough that the landscape changes
from year to year.

To study this field systematically, we build a unified corpus of
405~studies spanning 1973--2026 and read that corpus at two
levels.
The first stage is a field-level mapping that characterizes the field
over time and across a pipeline-centered taxonomy,
while positioning classical, ML-trained, and ML-prompted approaches
within the same map.
The second stage narrows the focus to ML-based studies and compares
their technical design patterns and the forms of evaluation evidence
they use to support claims about decompilation outputs.
Both stages rest on the same search, screening, and classification
process; what changes is the level of analysis.

We therefore address the following research questions:

\begin{description}[style=unboxed,leftmargin=0pt,nosep]
\item[RQ1.] How are classical and ML-based decompilation studies
  distributed over time and across the pipeline?
\item[RQ2.] How do ML-based decompilation studies vary in technical
  design and evaluation evidence?
\end{description}

RQ2 is operationalized through two sub-questions:

\begin{description}[style=unboxed,leftmargin=0pt,nosep]
\item[RQ2.1.] What technical design patterns do ML-based studies use to
  adapt ML to decompilation tasks?
\item[RQ2.2.] What forms of evaluation evidence do ML-based studies use
  to support claims of improvement in decompilation outputs?
\end{description}

Table~\ref{tab:rq-overview} summarizes where each question is addressed.

The analysis yields three main findings. First, classical and ML-based
decompilation are positioned differently across the pipeline rather than
forming a single replacement trajectory. Second, ML-based work is most
visible where recovered artifacts can be treated as prediction,
translation, or refinement targets, especially in symbol/name recovery,
output refinement, and End-to-End translation. Third, evaluation evidence
is highly task-specific, so reported improvements need to be interpreted
together with the target artifact, model-use strategy, and pipeline
position. The paper's main contribution is to make these relationships
explicit through a unified corpus, a pipeline-centered map, and a focused
comparison of ML-based technical designs and evaluation evidence.

\begin{table}[pos=t]
\centering
\caption{Research questions and where they are addressed in the paper.}
\label{tab:rq-overview}
\small
\begin{tabular}{p{0.10\columnwidth}p{0.33\columnwidth}p{0.43\columnwidth}}
\toprule
RQ & Main section & Focus of the answer \\
\midrule
RQ1 & Section~\ref{sec:landscape} &
Field-level mapping of decompilation studies over time and across the
pipeline-centered taxonomy, with classical and ML-based approaches
positioned within the same landscape. \\
RQ2 & Section~\ref{sec:focused-review} &
Focused review of how ML-based studies are characterized in terms of
technical design and evaluation evidence. \\
RQ2.1 & Section~\ref{sec:focused-review-model-strategies} &
Technical design patterns: architecture families, model-use strategies,
reusable backbones, and execution platforms. \\
RQ2.2 & Section~\ref{sec:focused-review-evaluation-metrics} &
Evaluation evidence: normalized metrics grouped by the kind of evidence
they provide for decompilation outputs. \\
\bottomrule
\end{tabular}
\end{table}

The remainder of this paper is organized as follows.
Section~\ref{sec:background} provides background on decompilation,
its history, and existing surveys.
Section~\ref{sec:methodology} defines the classification framework and
describes the methodology for corpus construction and the two-stage
analysis.
Section~\ref{sec:landscape} presents the field-level mapping results.
Section~\ref{sec:focused-review} provides the focused comparative
review of ML-based approaches.
Section~\ref{sec:discussion} interprets the cross-cutting implications
of the mapping and focused review,
Section~\ref{sec:threats} addresses threats to validity, and
Section~\ref{sec:conclusion} concludes the paper.

%% file: sections/2_background.tex
\section{Background and Related Work}\label{sec:background}

We first define decompilation and the scope of this survey, summarize
historical shifts in the role of decompiler-derived artifacts, then
introduce the pipeline view used to organize the field, and finally
position this survey relative to earlier reviews.

\subsection{Definition and Scope of Decompilation}\label{sec:bg-definition}

Decompilation is the process of recovering a human-understandable,
higher-level representation from a compiled low-level program.
Classical work describes this process as translating machine or assembly
code into a high-level language, or as attempting to invert compilation
from an executable program~\cite{Housel1974,Cifuentes1995b}.
In this survey, low-level programs include native binaries, bytecode,
and other executable representations.
The recovered output may be source-like pseudocode or source-level
program information, such as types and identifiers.

Decompilation is difficult because compilation discards or transforms
information that is explicit in source code.
Variable names, function names, comments, and much source-level type
information are often absent or only partially preserved in compiled
artifacts.
Structured conditionals and loops are lowered to branches, and
data-structure abstractions may appear only through memory accesses and
offset calculations.
Compiler optimizations such as inlining, loop unrolling, register
allocation, and instruction scheduling further obscure the relationship
between source and binary.
Consequently, decompilation does not generally recover the unique
original source program.
Instead, it reconstructs a higher-level representation from the
information that remains in the compiled artifact, aiming to preserve
program behavior and recover source-level structure where possible.

On this basis, our survey does not focus only on complete decompilers.
It also includes studies on individual processes that support
decompilation, such as transforming low-level code into analyzable
representations, recovering source-level information lost during
compilation, and improving or evaluating decompiler output.
This scope does not make the survey a review of binary analysis or
reverse engineering in general.
The detailed classification taxonomy and the operational inclusion and
exclusion criteria are defined in Section~\ref{sec:methodology}.

\subsection{Historical Shifts in Decompilation Artifacts}
\label{sec:bg-history}

\begin{figure*}[pos=t]
\centering
\includegraphics[width=0.92\textwidth]{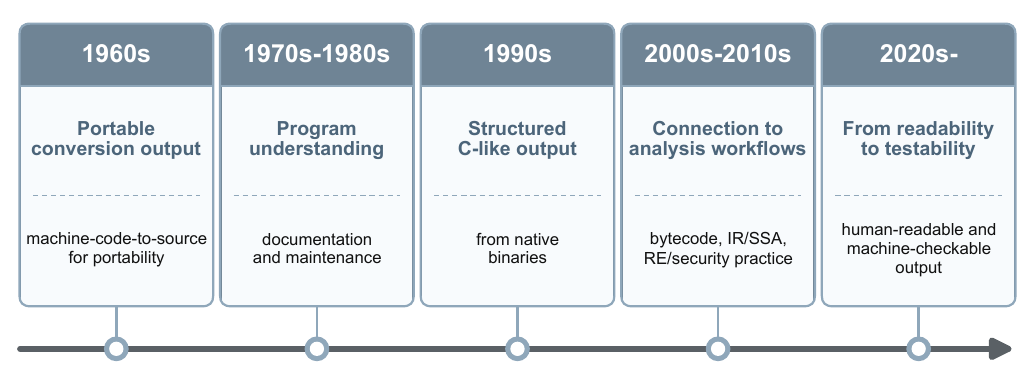}
\caption{Historical shifts in the role of decompiler-derived artifacts.
The labels summarize historically emphasized roles, not exhaustive or
mutually exclusive characterizations of each period.}
\label{fig:decompilation-history}
\end{figure*}

The role expected from decompiler-derived artifacts has changed over
time; Figure~\ref{fig:decompilation-history} summarizes these shifts.
Early work framed decompilation mainly as machine-code-to-source
conversion for portability and program migration, while later work also
used recovered representations for documentation, maintenance, and
program understanding~\cite{Housel1974,CifuentesPhD1994,Cifuentes1995b}.
As native-binary decompilation became systematized, structured C-like
output became the central human-facing artifact. In the 2000s and
2010s, decompilation increasingly interacted with bytecode, IR/SSA, and
reverse-engineering or security-analysis workflows. Recent work further
shifts attention from whether output is readable to whether it is
faithful, testable, and mechanically checkable~\cite{LiuWang2020,Dramko2024,LLM4Decompile2024}.

\subsection{Decompilation Pipeline}\label{sec:bg-pipeline}

Decompilation is not a single monolithic transformation.
Classical decompiler architectures group the process into
machine-dependent front-end processing, a machine- and
language-independent analysis module, and target-language-dependent
back-end code generation~\cite{CifuentesPhD1994}.
This view is useful even when modern studies do not implement the same
explicit pipeline.
Some studies focus on part of decompilation, such as recovering
low-level structure or reconstructing higher-level information.
Others improve the output of existing decompilers, evaluate the quality
of decompilation results, or generate higher-level representations
without explicitly separating the intermediate pipeline stages.
Although these studies differ in implementation, they all relate to the
process of recovering, improving, or evaluating source-like
representations from compiled artifacts.

We therefore use the pipeline view as an analytical framework, not as a
required tool architecture.
It allows us to locate studies by the part of the decompilation process
they primarily address.
Section~\ref{sec:framework} defines this framework in detail.

\subsection{Related Surveys and Reviews}\label{sec:bg-surveys}

Prior surveys and systematization studies provide important foundations
for understanding decompilation and its surrounding research areas.
Classical work established the conceptual basis of decompilation,
including its historical development, pipeline structure, and core
technical problems~\cite{CifuentesPhD1994,Cifuentes1995b}.
These works remain valuable because they define the problem space and
explain why recovering high-level programs from low-level code is
fundamentally difficult.
However, they predate the recent growth of ML-based and LLM-based
decompilation research, and therefore do not explain how learning-based
approaches are positioned within the broader decompilation pipeline.

A second line of surveys and SoK papers examines important components
or adjacent areas of binary analysis.
These studies clarify specific parts of the problem space, including
type inference on executables~\cite{Caballero2016}, binary
rewriting~\cite{Wenzl2019}, binary disassembly~\cite{Pang2021},
binary lifting~\cite{Liu2022}, and binary code
similarity~\cite{Haq2021}.
They are useful for understanding the assumptions and technical
difficulties of individual pipeline components and neighboring tasks.
However, their organizing scopes are component-specific or task-specific:
they do not treat decompilation itself as the object to be mapped across
the full pipeline.

More recent surveys address AI, ML, and LLMs in reverse engineering,
binary analysis, compilers, or decompilation.
These studies are valuable because they capture the rapid emergence of
learning-based techniques and summarize recent model-centered
developments~\cite{Ghimire2025,Hu2025,Zhang2026,Wang2026}.
However, most of them are organized around a technology family or a
broader reverse-engineering scope, and the closest survey, which
reviews LLM-based binary decompilation~\cite{Wang2026}, centers on
recent LLM-centered systems rather than on the field-wide relationship
between classical and ML-based decompilation.
Table~\ref{tab:related-surveys} summarizes how this survey relates to
these prior reviews.

\begin{table*}[pos=t]
\centering
\caption{Comparison with related surveys and SoK studies.
``Pipeline-wide'' indicates whether decompilation itself is mapped
across the full recovery pipeline; ``Classical+ML'' indicates whether
classical and ML-based approaches are compared within one frame.}
\label{tab:related-surveys}
\small
\begin{tabular}{lp{4.6cm}p{4.2cm}cc}
\toprule
Survey & Scope & Organizing frame & Pipeline-wide & Classical+ML \\
\midrule
Cifuentes~\cite{CifuentesPhD1994,Cifuentes1995b} &
  Native-binary decompilation & Decompiler architecture &
  \ding{51} & --- \\
Caballero and Lin~\cite{Caballero2016} &
  Type inference on executables & Component (types) & --- & --- \\
Wenzl et~al.~\cite{Wenzl2019} &
  Binary rewriting & Adjacent task & --- & --- \\
Pang et~al.~\cite{Pang2021} &
  Binary disassembly & Component (front-end) & --- & --- \\
Haq and Caballero~\cite{Haq2021} &
  Binary code similarity & Adjacent task & --- & \ding{51} \\
Liu et~al.~\cite{Liu2022} &
  Binary lifting & Component (IR) & --- & --- \\
Ghimire et~al.~\cite{Ghimire2025} &
  AI for reverse engineering & Technology family (AI) & --- & --- \\
Hu et~al.~\cite{Hu2025} &
  LLMs for reverse engineering & Technology family (LLM) & --- & --- \\
Zhang et~al.~\cite{Zhang2026} &
  LLMs and compilers & Technology family (LLM) & --- & --- \\
Wang et~al.~\cite{Wang2026} &
  LLM-based binary decompilation & Technology family (LLM) &
  --- & --- \\
\midrule
This survey &
  Decompilation, 405 studies (1973--2026) &
  Pipeline-centered taxonomy & \ding{51} & \ding{51} \\
\bottomrule
\end{tabular}
\end{table*}

This survey therefore takes a complementary view.
Rather than surveying a single component, adjacent task, or technology
family, we use the decompilation pipeline as the frame for comparing
classical and ML-based studies.
This perspective turns a fragmented literature into a structured map:
it shows which parts of the pipeline remain grounded in classical
analysis, where learning-based techniques have become visible, and where
reported progress depends on task-specific artifacts and evaluation
evidence.

%% file: sections/3_methodology.tex
\section{Methodology}\label{sec:methodology}\label{sec:framework}

\begin{figure*}[pos=t]
\centering
\includegraphics[width=0.95\textwidth]{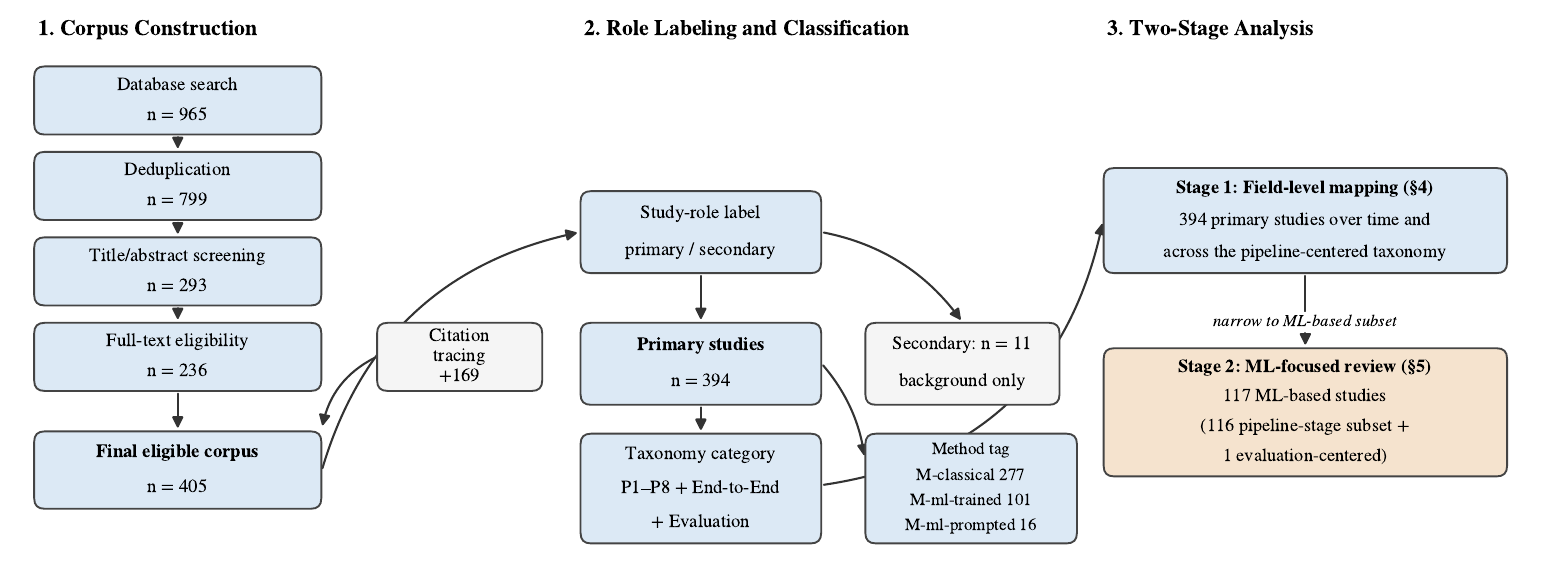}
\caption{Overview of the survey workflow: corpus construction, role
labeling and classification, and the two-stage analysis.}
\label{fig:methodology-flow}
\end{figure*}

This survey is organized as a two-stage study.
We first construct a systematic corpus of decompilation research and map
primary studies across a pipeline-centered classification framework.
We then narrow the analysis to ML-based primary studies and compare their
technical designs and evaluation evidence.
Both stages use the same search, screening, and classification workflow,
but differ in analysis scope and extraction detail.
Figure~\ref{fig:methodology-flow} summarizes the overall survey workflow,
from corpus construction to Stage~1 mapping and Stage~2 focused review.

\subsection{Overview of the Two-Stage Survey}
\label{sec:two-stage}

Our methodology builds one systematic corpus of decompilation research
and uses it for two connected analyses.
The first stage is a field-level mapping, informed by systematic mapping
guidance for software engineering~\cite{Petersen2008,Petersen2015}.
It describes the field over time and across the pipeline-centered
taxonomy, while positioning classical and ML-based approaches within the
same landscape (Section~\ref{sec:landscape}).
The second stage is a focused comparative review.
It examines where ML-based approaches have been applied and how they are
characterized in terms of technical design and evaluation evidence
(Section~\ref{sec:focused-review}).
We follow the guidelines of Kitchenham and Charters~\cite{Kitchenham2007}
for systematic reviews in software engineering and adopt the hybrid
database-plus-snowballing strategy recommended by
Wohlin et~al.~\cite{Wohlin2014,Wohlin2022}.

Each included study first receives a study-role label:

\begin{itemize}[nosep,leftmargin=*]
\item \textbf{\emph{Primary}}: papers whose main contribution is a novel
  method, tool, benchmark, dataset, empirical evaluation, comparative
  study, or other first-order technical artifact.
\item \textbf{\emph{Secondary}}: surveys, reviews, systematic literature
  reviews, systematizations of existing work, or conceptual papers that
  synthesize or contextualize the field without introducing a new
  technical artifact.
\end{itemize}

\emph{Secondary} studies are retained in the corpus for background
contextualization and search-completeness validation, but they are
excluded from the field-level mapping and focused review.

\emph{Primary} studies are then classified along two axes.
The first axis is the taxonomy category: one of the eight pipeline stages
(P1--P8) or one of two complementary categories, \emph{End-to-End} and
\emph{Evaluation}.
The second axis is the method tag.
We use three method tags in the corpus-level mapping:

\begin{itemize}[nosep,leftmargin=*]
\item \textbf{\emph{M-classical}}: methods whose central mechanism is not
  a learned model, including static and dynamic analysis,
  compiler-theoretic methods, formal methods, heuristics, and rule-based
  transformations.
\item \textbf{\emph{M-ml-trained}}: methods whose central mechanism is a
  machine-learning model whose parameters are fitted for the target
  task, whether trained from scratch on task-specific data or adapted
  from a reusable pre-trained backbone through continued pre-training or
  fine-tuning. This tag covers statistical learning, task-specific
  neural models, and fine-tuned language models.
\item \textbf{\emph{M-ml-prompted}}: methods that keep a (large) language
  model fixed and use it through prompting or in-context learning at
  inference time, including retrieval-augmented, analysis-augmented,
  feedback-guided, and tool-augmented prompting.
\end{itemize}

We refer to \emph{M-ml-trained} and \emph{M-ml-prompted} studies
collectively as \emph{ML-based} studies.
These tags serve two roles in the survey.
In Stage~1, they are used to position classical and ML-based work
within the same pipeline map.
In Stage~2, we focus on the ML-based primary studies.
Among them, studies assigned to P1--P8 or \emph{End-to-End} are analyzed
in detail because they propose or modify decompilation methods, either
at specific points in the pipeline or as cross-stage translation
systems.
Studies assigned to \emph{Evaluation} are kept in the analysis to
describe benchmarks, datasets, metrics, and comparative evaluation
practice, but they are not treated as direct method-comparison targets.

\subsection{Pipeline-Based Classification Framework}
\label{sec:taxonomy}

\begin{figure*}[pos=t]
\centering
\includegraphics[width=0.7\textwidth]{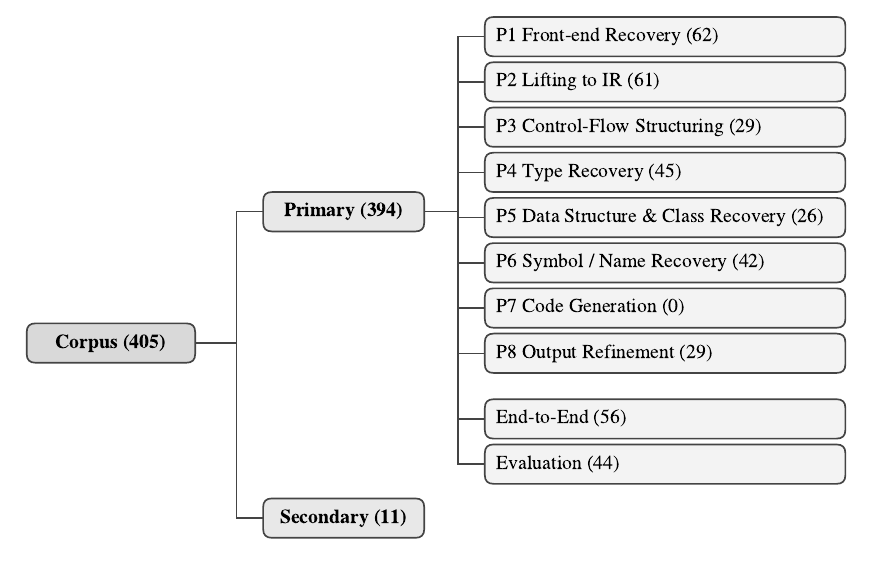}
\caption{Corpus structure.
Each primary study is assigned a primary taxonomy category for
corpus-level mapping.
P1--P8 denote pipeline stages; \emph{End-to-End} and \emph{Evaluation} are
complementary categories.
Each primary study additionally receives a primary method tag
(\emph{M-classical}, \emph{M-ml-trained}, or \emph{M-ml-prompted};
not shown).}
\label{fig:classification-tree}
\end{figure*}

\begin{table*}[pos=t]
\centering
\caption{Pipeline stage definitions (P1--P8).}
\label{tab:pipeline-stages}
\small
\begin{tabular}{clp{10.4cm}}
\toprule
Stage & Name & Description \\
\midrule
P1 & Front-end Recovery &
  Recovers instruction sequences, code regions, and function boundaries
  from raw binaries through disassembly, code discovery, and entry-point
  identification. \\
P2 & Lifting to IR &
  Translates architecture-specific machine instructions into a
  CPU-independent intermediate representation, such as LLVM~IR, VEX, or
  Ghidra P-code. \\
P3 & Control-Flow Structuring &
  Recovers high-level control structures (if/else, while, for, switch)
  from flat control-flow graphs. \\
P4 & Type Recovery &
  Infers primitive types, function signatures, calling conventions, and
  related type information for low-level programs. \\
P5 & Data Structure / Class Recovery &
  Reconstructs struct/class layouts, field boundaries, class
  hierarchies, and related object-structure information. \\
P6 & Symbol / Name Recovery &
  Recovers or predicts source-level identifiers, such as function names
  and variable names, for symbol-poor or decompiled programs. \\
P7 & Code Generation &
  Integrates recovered information and emits high-level source code
  (typically~C) from the internal representation. \\
P8 & Output Refinement &
  Improves the readability and quality of generated code through
  post-processing, including idiom recognition and LLM-based
  refinement. \\
\bottomrule
\end{tabular}
\end{table*}

Decompilation studies target different parts of the recovery process.
To compare classical and ML-based work without collapsing these different
tasks into a single category, we classify each primary study by the
dominant location of its technical contribution.
Our reference model builds on Cifuentes's classical decompiler
architecture~\cite{CifuentesPhD1994}.
This architecture provides two complementary levels of organization:
seven logical decompilation phases and three implementation modules.
The seven phases are syntax analysis, semantic analysis, intermediate
code generation, control-flow-graph generation, data-flow analysis,
control-flow analysis, and code generation.
These phases are grouped into a machine-dependent front-end, a machine-
and language-independent universal decompiling machine, and a
target-language-dependent back-end.
Thus, Cifuentes did not expand a three-category taxonomy into seven
categories; rather, the three modules group the seven logical phases.

We reorganize this classical basis into eight stages (P1--P8) for
survey classification.
The reorganization is needed because this survey classifies research
contributions rather than implementing a decompiler.
Several topics that were embedded in Cifuentes's semantic or data-flow
analysis, such as type recovery, data-structure recovery, and symbol/name
recovery, have become independent research areas with distinct methods
and evaluation practices.
We also add P8, Output Refinement, because recent studies increasingly
treat already generated decompiler output as the input artifact and
improve readability, recompilability, or source-like quality through
post-processing, rewriting, or ML-based refinement.

For corpus-level mapping, each primary study is assigned one primary
taxonomy category: one of the eight pipeline stages or one of two
complementary categories.
\emph{End-to-End} covers methods that formulate decompilation as a
cross-stage binary-to-source or low-level-to-source problem rather than
isolating one recovery stage as the main contribution.
\emph{Evaluation} covers primary studies whose main contribution is a
benchmark, dataset, metric, validation method, or comparative-evaluation
framework rather than a new recovery technique.
\emph{Secondary} studies, such as surveys, SoKs, and other
field-synthesizing papers, are not placed in the
\emph{Evaluation} category; they are handled through the study-role label
defined above.
Figure~\ref{fig:classification-tree} summarizes this corpus-level
classification structure.

When a study combines classical and ML-based techniques or spans multiple
pipeline stages, we assign the primary category and method tag according
to the study's main contribution and retain secondary roles as coding
notes for interpretation.
P1--P8 serve as an analytical reference model, not a claim about
implementation order.
Real decompilers exhibit cross-cutting analyses and iterative feedback
loops, so the taxonomy organizes the research landscape by dominant
contribution position; individual tools may traverse, combine, or bypass
these stages in different ways.

\subsection{Search Strategy}
\label{sec:search-strategy}

\begin{table}[pos=t]
\centering
\caption{Database search results (searched February~2026).}
\label{tab:db-hits}
\small
\begin{tabular}{lrrr}
\toprule
Database & Q1 & Q2 & Total \\
\midrule
ACM Digital Library & 221 & 19 & 240 \\
IEEE Xplore         & 253 & 11 & 264 \\
Web of Science      & 189 &  4 & 193 \\
arXiv               & 124 & 11 & 135 \\
Springer            & 116 &  0 & 116 \\
NDSS                &   8 &  1 &   9 \\
USENIX Security     &   8 &  0 &   8 \\
\midrule
\textbf{Total}      & \textbf{919} & \textbf{46} & \textbf{965} \\
\bottomrule
\end{tabular}
\end{table}

To capture both classical and ML-based decompilation research without
temporal bias, we impose no date restriction on the search.
We query five primary databases (ACM Digital Library, IEEE~Xplore,
Web of Science, Springer, and arXiv) and two supplementary security
venues (NDSS and USENIX Security) using two complementary query families
applied to titles and abstracts.

\paragraph*{Q1 (main query).}
\begin{list}{}{\leftmargin=0pt\rightmargin=0pt\topsep=2pt\parsep=0pt\itemsep=0pt}
\item\small\ttfamily\raggedright
decompil* OR "reverse compilation" OR "reverse compiler"
\end{list}

\noindent
The wildcard \texttt{decompil*} matches \emph{decompilation},
\emph{decompiler}, and \emph{decompiling}; the two phrase terms cover
direct synonyms used in early literature.
Because these terms are rarely used outside the binary-analysis context,
no additional AND~clause is needed.

\paragraph*{Q2 (complementary query).}
Phrases describing decompilation without using the root
\texttt{decompil*} are combined with a binary-context filter.
The full Boolean expression is:

\begin{list}{}{\leftmargin=0pt\rightmargin=0pt\topsep=2pt\parsep=0pt\itemsep=0pt}
\item\small\ttfamily\raggedright
("source code recovery" OR "source recovery" OR
"binary-to-source" OR "binary to source" OR
"binary lifting" OR "instruction lifting" OR
"lifting to IR" OR "lifting to LLVM" OR
"assembly-to-C" OR "assembly to C" OR
"program reconstruction")\\
AND\\
("binary code" OR "machine code" OR
"assembly code" OR executable OR disassembly OR
disassembler OR bytecode OR "binary analysis" OR
"stripped binary")
\end{list}

\noindent
Q2 supplements Q1 by recovering studies that address decompilation tasks
under alternative terminology.
The first group captures synonyms for decompilation; the second restricts
results to the binary-analysis context.
Table~\ref{tab:db-hits} reports the per-database hit counts.
The combined search yielded 965~records before deduplication.

\subsection{Study Selection}
\label{sec:study-selection}

Screening proceeds in three phases: deduplication, title/abstract
screening, and full-text eligibility assessment.
The inclusion and exclusion criteria are scope-based and independent of
the classification taxonomy defined in Section~\ref{sec:taxonomy},
thereby avoiding a self-fulfilling selection bias.

\paragraph*{Inclusion criteria.}
A study is included if it (IC1)~addresses decompilation or a constituent
sub-task of the decompilation pipeline,
(IC2)~makes a technical contribution (novel method, tool, benchmark,
dataset, or empirical evaluation), and
(IC3)~is written in English with full text accessible.

\paragraph*{Exclusion criteria.}
A study is excluded if it (EC1)~uses decompilation only as an
intermediate step for a downstream task without contributing to
decompilation itself,
(EC2)~targets out-of-scope subjects such as source-to-source translation
or hardware reverse engineering,
(EC3)~lacks sufficient technical content (e.g., position papers with no
evaluation),
(EC4)~is a duplicate version of an already-included study (the
peer-reviewed version is preferred over preprints), or
(EC5)~is a poster abstract, work-in-progress paper, master's thesis,
doctoral dissertation, technical report, or editorial.
Foundational dissertations may still be cited as contextual references,
but they are not counted as corpus studies.

\begin{figure}[pos=t]
\centering
\includegraphics[width=\columnwidth]{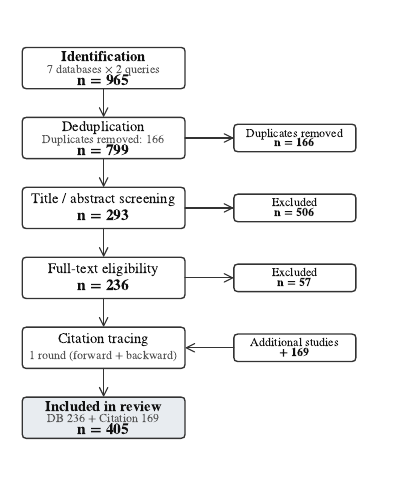}
\caption{Study selection flow. Counts report the final accounting after
eligibility reconciliation; the database search and citation tracing
(Section~\ref{sec:citation-tracing}) together yield the final corpus of
405~studies.}
\label{fig:selection-flow}
\end{figure}

Figure~\ref{fig:selection-flow} summarizes the selection process,
complementing the corpus-construction portion of
Figure~\ref{fig:methodology-flow}.
Initial deduplication removed 166~search records using bibliographic
metadata such as DOI and title, yielding 799~unique records.
Title and abstract screening excluded 506~records, leaving 293~records
for full-text eligibility assessment.
This later assessment included full-text review, publication-format and
publication-version checks, and scope verification; it excluded a further
57~records and produced 236~database-derived studies after eligibility
reconciliation.

\subsection{Citation Tracing}
\label{sec:citation-tracing}

To complement the database search, we performed one round of forward and
backward citation tracing using the eligible database-derived studies as
the seed set.
Decompilation research has developed across multiple research areas,
including compilers, program analysis, reverse engineering, and security;
therefore, even studies addressing the same or closely related problems
may differ in terminology and indexing.
Citation tracing was used to capture relevant studies that are difficult
to reach using search queries alone because of such terminology variation
or historical indexing differences.
This use of forward and backward citation tracing is consistent with
snowballing guidance for software engineering reviews~\cite{Wohlin2014}.

Candidate records obtained through citation tracing were deduplicated
against the database-derived set and screened using the same inclusion
and exclusion criteria.
After subsequent eligibility reconciliation, the database search
contributed 236~studies, and citation tracing contributed
169~additional studies.
The final corpus therefore contains \textbf{405~studies}
(236~database-derived $+$ 169~citation-derived).

\subsection{Classification and Data Extraction}
\label{sec:classification-extraction}

Data extraction is organized around the two stages of the survey.
For all included studies, we record bibliographic metadata and study
role; the study-role label separates primary studies from secondary
studies.
For primary studies, we additionally assign a taxonomy category and a
method tag.
The taxonomy category identifies the study's primary position in the
pipeline-based framework (P1--P8) or in the complementary
\emph{End-to-End} or \emph{Evaluation} categories.
Method tags are assigned only to primary studies and distinguish
\emph{M-classical}, \emph{M-ml-trained}, and \emph{M-ml-prompted} work
for the field-level mapping in Section~\ref{sec:landscape}.

Among primary studies, those tagged as \emph{M-ml-trained} or
\emph{M-ml-prompted} are carried forward to Stage~2.
For these studies, we first check whether the main contribution is
assigned to P1--P8, to \emph{End-to-End}, or to \emph{Evaluation}.
Studies assigned to P1--P8 or \emph{End-to-End} are analyzed in detail
in Section~\ref{sec:focused-review}, because they provide the main
basis for comparing ML-based decompilation methods.
Studies assigned to \emph{Evaluation} are also retained, but their role
is to describe evaluation infrastructure and practice, such as
benchmarks, datasets, metrics, and comparative evaluations.

\paragraph*{ML-focused extraction.}
For the ML-based studies analyzed in Stage~2, we extract a set of
additional attributes that support the comparative review in
Section~\ref{sec:focused-review}.
On the technical-design side, we record the model architecture family
(e.g., Transformer, RNN, GNN, or other specialized models), the primary
model-use strategy (from-scratch training, adaptation of a reusable
backbone, or prompting-based use), the canonical names of reusable or
external models when reported, and the execution platform targeted by
the study (native binaries, EVM, WebAssembly, JVM, or other VM
targets).
On the evaluation side, we record the reported metric names, normalize
them into a common vocabulary, and assign each metric to one of the
evaluation-evidence types defined in
Section~\ref{sec:focused-review-evaluation-metrics}; we also record
named datasets, benchmarks, and baselines as qualitative context.
When a study reports several architectures, strategies, or metrics, we
retain all of them in the extraction records but designate one primary
architecture family and one primary model-use strategy per study,
following the same dominant-contribution rule used for taxonomy
assignment.

\paragraph*{Quality assessment for the ML-focused review.}
After eligibility screening, we apply the Quality Assessment Criteria
(QAC) listed in Table~\ref{tab:qac-focused-llm} as part of data
extraction for the ML-focused review.
Quality assessment in software-engineering reviews can support not only
study selection, but also evidence synthesis and interpretation of
findings~\cite{Kitchenham2007,Yang2021}.
In this study, QAC is not used as an exclusion threshold.
Instead, it characterizes reporting completeness, methodological
transparency, and evidential support across the ML comparison subset,
following prior reviews that use quality assessment to interpret
evidence rather than to exclude studies by score~\cite{Tran2021,Lewowski2022}.

The assessment records whether each ML-based study clearly states its
objective and target problem, defines its main contribution, describes
the method and experimental setting with sufficient detail, provides
evidence adequate for its claims, reports analysis and results clearly,
and discusses limitations or threats to validity.
We do not aggregate these records into per-study scores or report them
as a separate scored analysis.
Instead, the per-study QAC records are retained in the replication
package and serve as interpretive context for the focused synthesis:
they underlie the cautions expressed in
Section~\ref{sec:focused-review} about partial reporting, such as the
incomplete reporting of reusable model names and evaluation settings.

\begin{table}[pos=t]
\centering
\caption{Quality Assessment Criteria for the ML-focused review.}
\label{tab:qac-focused-llm}
\small
\begin{tabular}{cp{5.8cm}}
\toprule
ID & Criterion \\
\midrule
QAC1 & Are the objectives and target problem clearly stated? \\
QAC2 & Is the main contribution clearly defined? \\
QAC3 & Is the research design described with sufficient detail for
        traceability? \\
QAC4 & Is the evidence adequate and sufficiently reported relative to
        the objectives? \\
QAC5 & Are the analysis and results clearly reported and do they
        support the conclusions? \\
QAC6 & Are limitations, scope, and validity threats discussed? \\
\bottomrule
\end{tabular}
\end{table}

\paragraph*{Auditability and reviewer process.}
Two reviewers participated in the survey process.
The first author performed the database search, screening,
classification, and data extraction.
A second reviewer independently checked the relevance judgments made
during the initial screening phase, verifying whether candidate records
fell within the decompilation scope defined by the inclusion and
exclusion criteria; disagreements were resolved through discussion until
consensus was reached.
Classification and extraction decisions beyond this screening check were
made by a single reviewer.
To keep these decisions auditable, we retain per-study records covering
the screening outcome, study-role label, taxonomy category, method tag,
and the ML-focused extraction attributes in the replication package,
so that individual decisions can be inspected and re-coded
independently.
The remaining risks of this reviewer design for classification and
extraction validity are discussed in Section~\ref{sec:threats}.

%% file: sections/5_landscape.tex
\section{The Decompilation Landscape}\label{sec:landscape}

This section answers \textbf{RQ1}: \emph{How are classical and ML-based
decompilation studies distributed over time and across the pipeline?}
The mapping yields a consistent picture.
Front-end and structural stages remain largely classical, ML-trained
work becomes most visible where source-level information must be
recovered, and prompted work is concentrated mainly in output-oriented
settings.
The rest of the section first situates the corpus as a whole
(Section~\ref{sec:landscape-overview}), then profiles each stage and
complementary category in the heatmap
(Section~\ref{sec:landscape-stages}), and finally summarizes the
mapping findings while explaining why the map should not be read as a
performance comparison (Section~\ref{sec:landscape-summary}).
The stage profiles keep ML discussion at the level of where learning
enters the map; detailed comparisons of ML-based techniques and
evaluation evidence are deferred to Section~\ref{sec:focused-review}.

%% ─────────────────────────────────────────────────────────────────────
\subsection{Field-Wide Structure}\label{sec:landscape-overview}

\begin{figure}[pos=t]
  \centering
  \includegraphics[width=0.72\columnwidth]{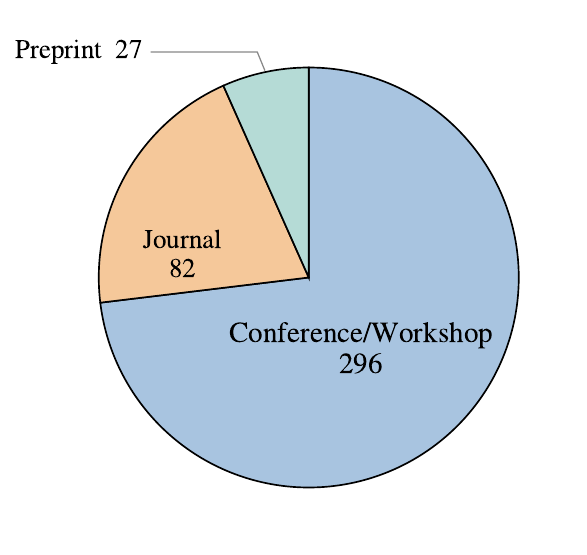}
  \caption{Publication-type distribution of the 405 included studies.
    Conferences and workshops account for 73\% of the corpus.}
  \label{fig:venue-dist}
\end{figure}

\begin{figure*}[pos=t]
  \centering
  \includegraphics[width=\textwidth]{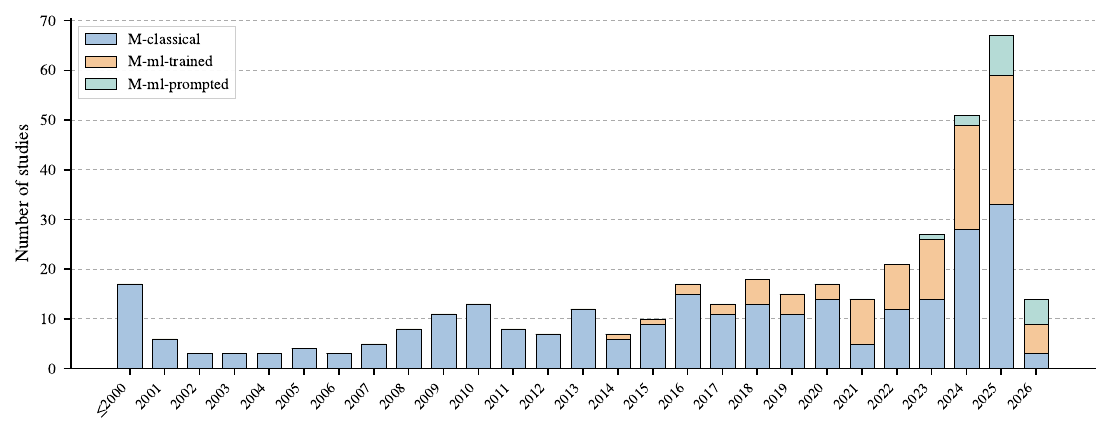}
  \caption{Publication-year distribution of the 394 primary studies with publication years,
    stratified by method tag.
    Classical studies dominate throughout, but ML-based work (first
    trained models, then prompted LLMs) accelerates sharply after 2020.}
  \label{fig:year-trend}
\end{figure*}

The final corpus comprises 405~studies, of which 394 are primary
studies and 11 are secondary studies (surveys, SoKs, and reviews)
retained for background reference but excluded from the mapping
analysis.
Among the 394 primary studies, 277 are classified as classical,
101 as ML-based with trained models, and 16 as ML-based with prompted
or in-context-learning approaches.

Figure~\ref{fig:venue-dist} shows the publication-type breakdown of the
full 405-study corpus.
Conferences and workshops account for 296~studies (73\%), journals for
82 (20\%), and preprints or other informal publications for 27 (7\%).
Table~\ref{tab:top-venues} lists the most frequent individual venues
and publication outlets.
Beyond arXiv, security conferences form the largest named group:
USENIX Security, IEEE S\&P, CCS, and NDSS together account for
54~studies, reflecting the fact that malware analysis and vulnerability
discovery are the primary application drivers.
Reverse-engineering venues (WCRE/SANER) contribute 18~studies.
PL-oriented venues (PLDI, CC, PACMPL) and SE venues (ASE, ISSTA) each
contribute a steady share, showing that the field sits at the
intersection of security, programming languages, and software
engineering.
The 27~preprints are concentrated in the post-2022 ML wave, consistent
with the rapid publication cycle of LLM-related research.

Figure~\ref{fig:year-trend} shows the publication-year distribution.
Publication volume was modest and steady from the 1970s through the
mid-2000s, reflecting decompilation's origins as a niche concern in
program analysis and binary translation.
A first acceleration occurred around 2010, alongside growing tool
availability and increasing interest in binary security.
A sharper inflection appeared after 2020, when ML-based
approaches, first task-trained neural models and then prompted large
language models, entered the field in significant numbers.
This acceleration is not uniform across the pipeline: it is concentrated
in stages where the task can be framed as prediction from learned
patterns (P6, P8, and End-to-End), while stages dominated by formal
reasoning (P2, P3) remain almost entirely classical.

\input{tables/table_top_venues}

In this mapping, the pipeline-centered taxonomy comprises the P1--P8
pipeline stages plus two complementary categories: End-to-End and
Evaluation.
End-to-End studies, which treat binary-to-source translation as a
cross-stage problem rather than assigning the contribution to a single
pipeline stage, account for 56~papers.
Evaluation studies, whose primary contribution is methodological rather
than a new recovery step, account for 44~papers.
Section~\ref{sec:landscape-stages} discusses both categories separately
from the P1--P8 sequence, and Section~\ref{sec:focused-review} later
examines the ML-based subset of End-to-End and other cross-stage
systems.

%% ─────────────────────────────────────────────────────────────────────
\subsection{Pipeline-Stage Profiles}\label{sec:landscape-stages}

\begin{figure*}[t]
  \centering
  \includegraphics[width=\textwidth]{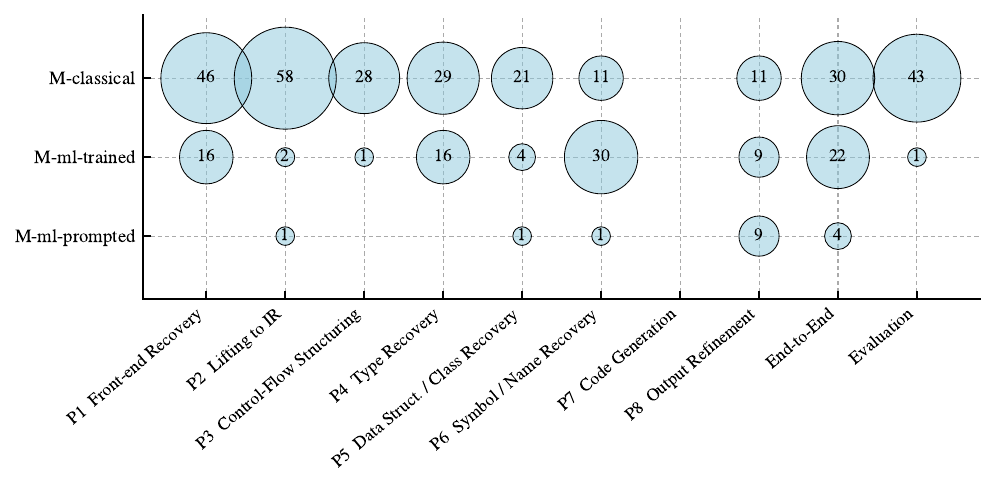}
  \caption{Distribution of primary studies across pipeline stages,
    complementary categories, and method categories.
    Classical work concentrates in front-end and structural stages
    (P1--P3), while ML-based studies dominate symbol recovery (P6) and
    output refinement (P8).
    The matrix covers the 394 primary studies in the mapping corpus.}
  \label{fig:heatmap}
\end{figure*}

Figure~\ref{fig:heatmap} is the main evidence for this subsection.
Rather than treating each pipeline stage as a separate technical
history, we read the matrix as a stage-by-method profile: for each
stage, the relevant question is how classical, ML-trained, and
ML-prompted studies are distributed, and what that distribution suggests
about the role of the stage in the broader decompilation landscape.
The counts in the heatmap are therefore used as mapping evidence, not as
a performance comparison or a measure of technical maturity.

The matrix shows a markedly uneven distribution.
Classical work remains concentrated in the upstream and structural
parts of the reference model, especially front-end recovery, lifting to
IR, and control-flow structuring.
ML-trained work appears more strongly in stages associated with
source-level information, most visibly type recovery and symbol/name
recovery.
Prompted work is even more concentrated: it is sparse across most
pipeline stages but visible in output refinement, a stage whose inputs
and outputs are often closer to source-like text.
End-to-End studies are kept as a complementary cross-stage category
rather than decomposed into P1--P8, because they formulate
binary-to-source recovery as an integrated translation problem.

Viewed at this level, the heatmap does not show a uniform shift from
classical decompilation to ML-based decompilation.
It shows a positional pattern.
Classical studies occupy the stages where the recovered artifact is
closest to the binary and where correctness is constrained by instruction
semantics, control-flow consistency, or compiler-style invariants.
ML-based studies become more visible as the target artifact moves toward
information that compilation, stripping, or optimization has removed:
types, identifiers, source-like rewrites, and whole-function
translations.
This pattern is especially clear when P1--P3 are compared with P6, P8,
and End-to-End.
The former stages provide much of the structural substrate on which
decompilation depends, whereas the latter stages are closer to semantic
completion and source-level rendering.
P4 and P5 sit between these poles: they still rely on constraints from
program analysis and memory layout, but they also expose prediction
problems in which learned models can recover or rank missing source-level
information.

The distinction between ML-trained and prompted studies sharpens this
reading.
ML-trained work is visible in both middle and downstream stages, because
supervised prediction can be adapted to type labels, function names,
variable names, and translation-like tasks.
Prompted LLM work, by contrast, appears mainly where the input and output
can be represented as source-like text or as an integrated generation
problem.
Its concentration in output refinement and End-to-End studies therefore
reflects not just recency, but also the kinds of artifacts for which
prompting is currently easiest to operationalize.
The heatmap therefore positions ML less as a replacement for the
pipeline than as a set of learned components concentrated around
source-level completion and output-oriented recovery.

The stage profiles below therefore follow a fixed structure.
Each stage starts from the heatmap count, briefly identifies the
classical baseline that anchors the stage, notes the task form in which
ML-based work appears or remains sparse, and offers a cautious reading
of the distribution.
Detailed comparisons of ML models, benchmarks, metrics, and reported
improvements are deferred to Section~\ref{sec:focused-review}.

\paragraph*{P1: Front-end Recovery.}
Figure~\ref{fig:heatmap} counts 46 classical studies, 16 ML-trained
studies, and no prompted studies in front-end recovery.
This distribution makes P1 a classical-dominant stage, with classical
analysis providing the front-end artifacts on which later decompilation
depends.
Classical studies establish these artifacts through instruction
recovery, function-boundary detection, early control-flow recovery, and
jump-table resolution \cite{Cifuentes2001b,Meng2016,AlvesFoss2019}.

The ML-based task form in P1 is mainly prediction over ambiguous
front-end decisions in binaries, especially boundaries, entry points,
code/data labels, and local control-transfer targets.
Examples include function-boundary prediction and related boundary or
labeling tasks, as
illustrated by early neural function recognition~\cite{Shin2015}.
Together, these studies suggest that learning enters P1 where front-end
ambiguities can be formulated as supervised prediction, while the
classical front end remains the main basis for producing the artifacts
that later pipeline stages consume.

\paragraph*{P2: Lifting to IR.}
Figure~\ref{fig:heatmap} counts 58 classical studies, two ML-trained
studies, and one prompted study in lifting to intermediate
representations.
Among the pipeline stages, P2 is one of the most strongly anchored in
classical work.
Technically, this stage is less about inferring source-like information
than about constructing semantics-preserving representations that later
analyses can trust.
Classical work in this stage focuses on moving low-level code into
analyzable forms, including SSA-based decompiler IRs, compiler-level IR
lifting, architecture-neutral representations, and proof-carrying or
verified lifting
\cite{REcompile2013,Anand2013,Verbeek2022}.
The clearest ML-trained example appears in a specific translation
setting: learned support for instruction- or assembly-to-IR translation,
rather than a broad replacement of lifter design
\cite{Hasabnis2016}.
In the corpus, then, P2 remains centered on classical methods; the few
learning-based entries are too sparse and heterogeneous to define a
broad alternative to lifter design.

\paragraph*{P3: Control-Flow Structuring.}
Figure~\ref{fig:heatmap} counts 28 classical studies, one ML-trained
study, and no prompted studies in control-flow structuring.
P3 is similarly dominated by classical work.
Classical work in this stage addresses the recovery of high-level
control constructs from flat control-flow graphs, including structured
conditionals and loops, goto elimination, semantics-preserving
transformations, and source-fidelity-oriented structuring
\cite{Cifuentes1996b,NoMoreGotos2015,SAILR2024}.
This line of work gives P3 a different profile from stages such as type
or name recovery: the central artifact is not missing source-level
information, but a control-flow structure whose readability and source
fidelity must be balanced against consistency with the recovered CFG.

The single ML-based entry in P3 takes a localized form: rather than
replacing structuring algorithms broadly, it appears as classification
support.
It formulates ambiguity in Java bytecode control flow as a local
statement-classification problem, using learned classification to guide
a Java decompiler~\cite{Sateanpattanakul2022}.
P3 thus remains grounded in classical analyses of control-flow
consistency, goto placement, and source fidelity; in the current corpus,
ML-based work appears only through the narrow task form of control-flow
classification.

\paragraph*{P4: Type Recovery.}
Figure~\ref{fig:heatmap} counts 29 classical studies, 16 ML-trained
studies, and no prompted studies in type recovery.
Unlike P2 and P3, P4 is not only a classical-heavy stage: the heatmap
also shows a substantial ML-trained presence.
Classical work in this stage treats type recovery as inference over
evidence left in compiled artifacts, including type constraints,
abstract interpretation, type propagation, and
calling-convention-related evidence for type and function-signature
recovery.
This line ranges from early type-based decompilation to principled
reverse engineering of binary types, polymorphic machine-code type
inference, and recent combined-constraint signature recovery
\cite{Mycroft1999,TIE2011,Retypd2016,Huang2025}.

The ML-trained presence in P4 centers on prediction or recovery of
source-level type information erased during compilation.
Unlike P3, where ML appears only as a narrow control-flow
classification aid, P4 contains a broader set of learned formulations:
variable and primitive-type prediction, function-signature prediction,
and type recovery from instruction sequences, data-flow graphs,
intermediate representations, or decompiler artifacts.
Examples include neural function-signature inference, fine-grained type
recovery, and graph-based type inference for stripped binaries
\cite{Chua2017,StateFormer2021,Zhu2024}.
This profile places P4 closer to the ML-heavy source-information stages
than P2 or P3.
Type recovery still depends on constraints and program-analysis
evidence, but ML-trained work becomes visible where that evidence can
be used to predict type labels, function signatures, or other erased
type information.

\paragraph*{P5: Data Structure / Class Recovery.}
Figure~\ref{fig:heatmap} counts 21 classical studies, four ML-trained
studies, and one prompted study in data-structure and class recovery.
Like type recovery, P5 recovers information lost or obscured during
compilation, but its heatmap profile is more classical-leaning than P4.
Classical work in this stage focuses on composite memory and object
structures rather than individual type labels: struct and array layouts,
field boundaries, C++ class hierarchies, vtables, and dynamic pointer
structures.
Representative work recovers data structures from memory-use evidence,
reconstructs class hierarchies from object-oriented binary artifacts,
or models variable and structure recovery under compilation-induced
uncertainty \cite{Howard2011,Pawlowski2017,OSPREY2021}.

The smaller ML-based presence in P5 does not suggest a broad learned
replacement for structural recovery.
Instead, ML-based work appears as support for specific subtasks, such as
statistical class-hierarchy reconstruction, stack-frame size prediction,
GNN-based inference of structural semantics, and fine-tuned LLM-based
recovery of user-defined structures and their symbols
\cite{Katz2018b,StackBERT2021,Sakamoto2025,ReSym2024}.
The prompted entry fits a related hybrid pattern, where LLM reasoning is
coupled with static-analysis evidence for structure recovery
\cite{HyRES2026}.
P5 therefore remains grounded in memory-layout and object-structure
evidence; ML-based work enters where structural hypotheses can be
ranked, layout properties can be predicted, or recovered structures and
symbols can be checked with learned or hybrid support.

\paragraph*{P6: Symbol / Name Recovery.}
Figure~\ref{fig:heatmap} counts 11 classical studies, 30 ML-trained
studies, and one prompted study in symbol and name recovery.
Within the pipeline stages, P6 is the clearest concentration of
ML-trained work.
Classical work in this stage is concentrated around name recovery from
explicit, matched, or domain-specific evidence:
library-function identification through signatures or semantic
descriptors, identifier deobfuscation, and framework-specific symbol
recovery.
These studies provide an important baseline, but their scope is mainly
known-function or special-case recovery rather than general naming of
arbitrary stripped functions or decompiler-introduced variables
\cite{Jacobson2011,Cimato2005,Wen2023}.

The ML-trained profile of P6 is therefore best read as prediction or
generation of lost source-level identifiers.
The task appears in several related forms: function-name prediction
from stripped binary bodies or analysis-derived representations,
variable-name recovery from decompiler output, decompiled identifier
renaming, and broader recovery of symbol/debug information.
Examples range from structured prediction of debug information to
extreme multi-label function naming, context-sensitive function-symbol
prediction, and transformer or generative-model approaches to variable
renaming
\cite{Debin2018,PatrickEvans2023,SymLM2022,Dramko2023,Pal2024,Xu2025}.
The single prompted entry is narrower: it addresses firmware-specific
function renaming with an LLM, so the P6 profile remains driven mainly
by ML-trained studies~\cite{Liu2025c}.
P6 is thus where ML-trained work most visibly enters the pipeline as
source-level label prediction: compilation and stripping remove
human-readable names, while binaries, decompiler artifacts, call
context, and program-analysis features still provide signals from which
plausible names can be learned.

\paragraph*{P7: Code Generation.}
Figure~\ref{fig:heatmap} assigns no primary studies to code generation
under any method tag.
This absence should not be read as the absence of code emission in
decompilers.
IR-to-source emission is an operationally necessary part of many
decompilation systems, but in the mapped corpus it is not isolated as
the primary research contribution of any included study.
Instead, code generation is usually embedded inside complete
decompilers, end-to-end binary-to-source systems, or output-refinement
work.
Thus, P7 appears in the heatmap as a gap not because the function is
unimportant, but because it is not commonly separated as an independent
research category.

\paragraph*{P8: Output Refinement.}
Figure~\ref{fig:heatmap} counts 11 classical studies, nine ML-trained
studies, and nine prompted studies in output refinement.
P8 is therefore not dominated by a single method family in the same
way as P6; instead, it is the pipeline stage where prompted work is
most visible.
Classical work in this stage treats decompiler output as an artifact
that can be improved after generation.
Its tasks include readability-oriented rewriting, meta-decompilation,
expression or obfuscation simplification, language-specific rewriting,
and recompilation-oriented workflows.
Examples include usability-oriented decompiler transformations for
malware analysis, meta-decompilation across Java decompilers,
MBA simplification, non-C rule-based output improvement, and
partially recompilable decompilation
\cite{DREAM2016,Harrand2020,MBASED2024,ideco2025,Reiter2025}.

The ML-based task form in P8 is centered on rewriting, repairing, or
augmenting source-like decompiler output rather than recovering a
low-level pipeline artifact.
Prompted systems are especially visible here: they use LLMs to refine
pseudocode, correct distorted structures, improve recompilability, or
translate decompiler output toward compilable and executable code
\cite{DeGPT2024,DecLLM2025,PseudoFix2025,PCodeTrans2026}.
ML-trained work appears in adjacent forms, including learned
augmentation of names and types, fine-tuned decompiler-backend or local
decompilation improvement, and multi-task reverse-engineering
assistance built on decompiler artifacts
\cite{DIRTY2022,DLiFT2025,SK2Decompile2025,ReCopilot2025}.
In P8, source-like artifacts make ML-based rewriting and repair
particularly visible in the corpus; how readability, recompilability,
and semantic faithfulness trade off against each other is a question for
the focused comparison, not one the heatmap alone can settle.

\paragraph*{End-to-End.}
Figure~\ref{fig:heatmap} counts 30 classical studies, 22 ML-trained
studies, and four prompted studies in the End-to-End category.
Unlike P1--P8, this row is not a pipeline stage.
It collects studies that formulate decompilation as a cross-stage
binary-to-source or low-level-to-source problem, rather than isolating a
single recovery step as the primary contribution.
Classical work in this category includes whole decompiler designs,
machine-code-to-source translation, retargetable decompilation, and
systems that connect recovered source code to analysis, debugging, or
validation workflows
\cite{Housel1974,Cifuentes1995b,Cifuentes1998a,Verbeek2020}.
The End-to-End row is thus a complementary formulation: it shows where
studies treat the pipeline as an integrated recovery problem, not where
a new terminal stage is added after P8.

The ML-based task form in End-to-End work is function-level or
program-level generation from low-level artifacts.
Early neural studies framed this as sequence or tree-structured
translation from assembly or binary-derived representations to source
code, and later work expands the same formulation with Transformer
models, fine-tuned LLMs, type-inference support, retrieved context,
program-analysis artifacts, or feedback-guided prompting
\cite{Katz2018a,Coda2019,SLaDe2024,LLM4Decompile2024}.
The four prompted studies make LLM-based cross-stage generation visible
in this row, but the profile is still driven mainly by ML-trained
systems rather than by prompted-only use.
This distribution suggests that End-to-End work is where ML most
directly recasts decompilation as translation or generation, while the
pipeline map remains necessary for interpreting what such systems
compress, bypass, or reintroduce as auxiliary analysis.

\paragraph*{Evaluation.}
Figure~\ref{fig:heatmap} counts 43 classical studies, one ML-trained
study, and no prompted studies in the Evaluation category.
Like End-to-End, Evaluation is a complementary category rather than a
pipeline stage.
Its heatmap profile is therefore not evidence that evaluation itself is
a classical recovery step.
Instead, it indicates that evaluation-centered primary studies in the
corpus are mostly organized around non-learning contributions such as
benchmarks, tool comparisons, testing frameworks, validation methods,
human studies, and metrics.
This reading leads directly to the comparability caution in the summary:
the heatmap shows where studies are located, but evaluation evidence
remains stage- and task-specific.

%% ─────────────────────────────────────────────────────────────────────
\subsection{Summary of Mapping Findings for RQ1}\label{sec:landscape-summary}

In answer to RQ1, classical and ML-based decompilation studies are
distributed unevenly over time and across the pipeline, and this
unevenness has become more visible with the recent growth of ML-based
work.
At the corpus level, the primary-study map is dominated historically by
classical work, while End-to-End and Evaluation appear as complementary
categories that capture cross-stage translation and methodological
contributions rather than additional pipeline stages.
At the pipeline level, the central finding is positional rather than
competitive: classical and ML-based approaches occupy different parts of
the map.
Front-end recovery, IR lifting, and control-flow structuring remain
dominated by classical methods grounded in compiler theory, static
analysis, and formal reasoning.
Type recovery is the clearest mixed region, where classical
constraints and learned prediction both have substantial presence,
whereas data-structure and class recovery remains more classical-leaning
but includes targeted ML and hybrid entries.
Symbol recovery and output refinement show the strongest ML presence
among pipeline stages, while end-to-end translation has grown as a
separate cross-stage formulation.
Evaluation is likewise complementary to the pipeline: it records
benchmarking, tool comparison, validation, and measurement contributions
rather than a separate recovery stage.

This positioning gives a more precise reading of where ML sits in the
field.
ML-based work is not spread evenly across all decompilation tasks.
It concentrates where the output must recover or generate source-level
information that is no longer directly present in the binary: plausible
types, names, readable code, repaired pseudocode, or source-like
translations.
ML-trained studies cover a broader range of these prediction and
generation tasks, whereas prompted LLM studies are still concentrated
mainly in output-oriented and End-to-End settings.
Conversely, the upstream and structural stages remain classical
territory, not because they are unimportant for ML-based systems, but
because they provide the constrained artifacts and invariants that many
downstream learned components still depend on.

The temporal pattern reinforces this structural picture.
Publication volume is relatively steady from the 1970s through the
mid-2000s, grows around 2010, and changes most sharply after 2020 with
the rise of ML-trained and then prompted studies.
Classical studies dominate the historical corpus, and the upstream
stages still rest on them.
The post-2020 ML wave has therefore not replaced the classical
decompilation pipeline uniformly; rather, the map is consistent with the
interpretation that learning has concentrated most visibly where recovery
requires plausible source-level completion in addition to exact
structural inversion.

These findings are mapping evidence, not a performance ranking.
A stage with many studies is not necessarily more solved, and a stage
with many ML-based studies is not necessarily easier or more mature.
The target artifacts, benchmarks, metrics, and oracles differ across
stages, so the heatmap does not provide a common scale for comparing
progress across the pipeline.
Section~\ref{sec:focused-review} therefore moves from map-level
distribution to evidence-level comparison: it narrows the analysis to
ML-based studies and reads their technical design patterns and
evaluation evidence within the mapped landscape.

%% file: tables/table_top_venues.tex
\begin{table}[pos=t]
\centering
\caption{Most frequent publication venues and outlets (5+ studies).}
\label{tab:top-venues}
\small
\begin{tabular}{lr}
\toprule
Venue / outlet & Studies \\
\midrule
arXiv                      & 27 \\
USENIX Security            & 18 \\
WCRE / SANER               & 18 \\
IEEE S\&P                  & 14 \\
NDSS                       & 12 \\
CCS                        & 10 \\
ASE                        &  9 \\
ISSTA                      &  9 \\
PACMPL                     &  9 \\
BAR                        &  7 \\
IEEE TSE                   &  7 \\
PLDI                       &  7 \\
TOSEM                      &  7 \\
CC                         &  6 \\
PACMSE                     &  6 \\
AsiaCCS                    &  5 \\
ASPLOS                     &  5 \\
DSN                        &  5 \\
ICPC                       &  5 \\
\bottomrule
\end{tabular}
\end{table}

%% file: sections/6_focused_review.tex
\section{ML-Based Decompilation: A Comparative Review}\label{sec:focused-review}

This section answers \textbf{RQ2}: \emph{How do ML-based decompilation
studies vary in technical design and evaluation evidence?}
Section~\ref{sec:landscape} located classical, ML-trained, and prompted
studies on a common pipeline-centered map.
This section narrows that map to the ML-based subset and asks a different
kind of question: not whether ML is globally better than classical
decompilation, but how ML-based studies are designed and what kinds of
evidence they use when claiming improvements in decompilation outputs.
The review therefore treats the pipeline map as context, while organizing
the detailed synthesis around two evidence axes: technical design
patterns (RQ2.1) and evaluation evidence (RQ2.2).
This framing avoids forcing heterogeneous studies into a single
leaderboard and instead makes explicit which claims are supported by
which kinds of extracted evidence.

%% ─────────────────────────────────────────────────────────────────────
\subsection{Scope of the ML-Focused Review}\label{sec:focused-review-scope}

The ML-focused review covers the 117 ML-based primary studies identified
in the mapping study: 101 studies using trained models and 16 using
prompted or in-context-learning approaches.
Among them, 116 propose or modify decompilation methods located in the
pipeline map, that is, in P1--P8 or \emph{End-to-End}, and form the
harmonized comparison subset used for the technical-design and
evaluation-evidence analyses in this section.
The remaining study is evaluation-centered rather than pipeline-stage
centered, and is retained as contextual evidence about evaluation
infrastructure.
This scope follows directly from RQ1: the mapping first shows where
ML-based work sits relative to the broader decompilation literature, and
the focused review then characterizes the ML-based studies located within
that map.

The ML subset is a recent and methodologically mixed body of work,
though year and venue distributions serve here as background context,
not as separate evidence axes for RQ2.
Task-trained neural models appear before the current LLM wave, while
prompted systems enter the corpus only after the spread of large
code-oriented foundation models.
The venue distribution is likewise mixed: security venues remain a
visible home for work on stripped binaries, obfuscation, and semantic
correctness, while software-engineering venues and preprints capture a
growing connection to code intelligence and foundation-model reuse.
This mixture is examined through the two sub-questions defined in
Section~\ref{sec:introduction}:
\begin{description}[style=unboxed,leftmargin=0pt,nosep]
\item[RQ2.1.] What technical design patterns do ML-based studies use to
adapt ML to decompilation tasks?
\item[RQ2.2.] What forms of evaluation evidence do ML-based studies use
to support claims of improvement in decompilation outputs?
\end{description}

The remainder of this section does not repeat the stage-by-stage reading
of Section~\ref{sec:landscape-stages}.
Nevertheless, pipeline context remains useful because technical designs
and evaluation oracles are shaped by the artifact being recovered.
The main synthesis remains organized by RQ2.1 and RQ2.2, with pipeline
stages used only as context when a design or evaluation pattern is
stage-sensitive.

\subsection{Technical Design Patterns (RQ2.1)}
\label{sec:focused-review-model-strategies}

This subsection answers RQ2.1 by examining how ML-based studies turn
machine-learning models into decompilation methods.
Model architecture is only one part of that design: the same broad model
family can be trained from scratch, adapted from a code-oriented
backbone, or used through prompting, and these choices often correspond
to different target artifacts and execution environments.
We therefore read the ML-focused subset through four design dimensions:
architecture families, model-use strategies, reusable model backbones,
and execution platforms.
These dimensions are review aids rather than a new taxonomy; pipeline
context is introduced only where it helps explain why a design choice
appears in a particular task setting.

\subsubsection*{Architecture families}
\begin{figure}[pos=t]
  \centering
  \includegraphics[width=\columnwidth]{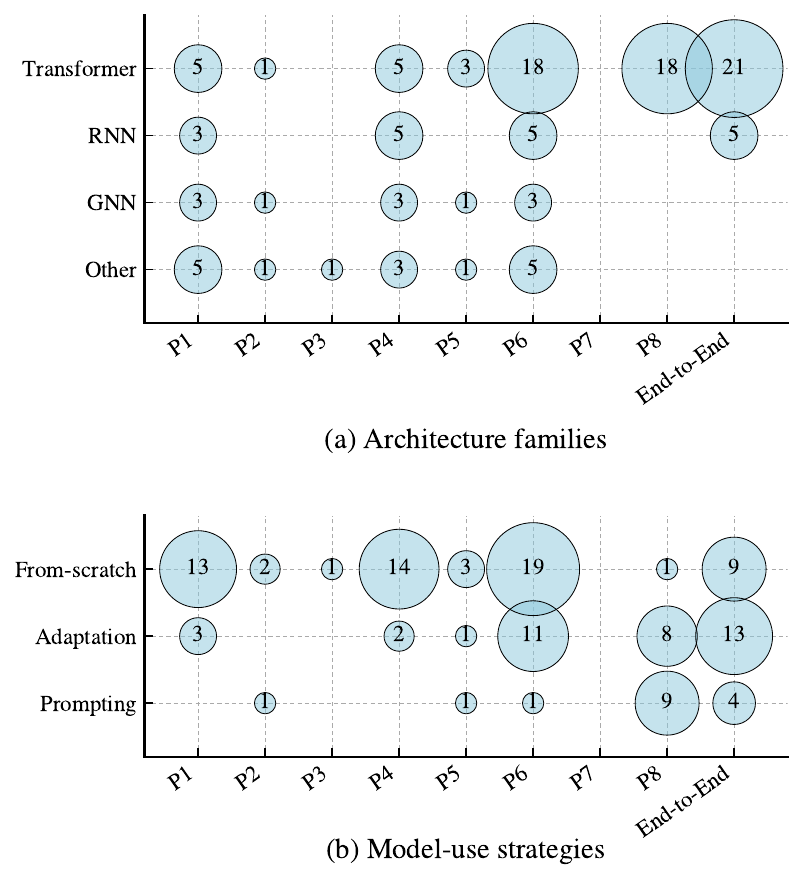}
  \caption{Distribution of primary model architecture families
  and primary model-use strategies across pipeline stages in the
  116-study ML comparison subset. Panel~(a) groups low-frequency
  specialized architecture families under \textit{Other};
  Panel~(b) groups analysis-, feedback-, and context-augmented
  prompting under \textit{Prompting}.}
  \label{fig:ml-technical-profile}
\end{figure}

The extraction records contain 125 architecture-family labels across the
116-study ML comparison subset, with Transformers accounting for 71 of
these labels.
Figure~\ref{fig:ml-technical-profile} shows that the same concentration
remains visible in the primary-family view across pipeline stages.
This does not mean that one architecture has uniformly replaced the
rest across the whole pipeline.
Instead, the Transformer concentration is strongest in stages where the
task operates over lexical, pseudocode, or source-like artifacts: P6,
P8, and End-to-End each show a large Transformer primary-family presence.
In these settings, Transformer models are used for identifier recovery,
pseudocode augmentation, decompiler-output refinement, and function-level
code generation.
Representative examples include Transformer-based identifier renaming,
unified pre-training over stripped-binary pseudocode, compact
assembly-to-C generation, and larger LLM-based decompilation systems
\cite{DIRECT2021,HexT52023,SLaDe2024,LLM4Decompile2024}.
Thus, the concentration of Transformers is best read not simply as a
model trend, but as consistent with the view that reusable sequence and
language-model architectures are especially visible when the task is to
predict, generate, or rewrite source-like artifacts.

The non-Transformer families are smaller, but they are not merely
residual categories.
They often correspond to task forms where local sequence structure,
graph-structured program evidence, or specialized symbolic models remain
natural.
RNNs appear in early function-boundary recognition and earlier
sequence-to-sequence decompilation tasks~\cite{Shin2015,Katz2018a,Coda2019};
GNNs appear where the method explicitly constructs instruction-flow,
data-flow, or variable graphs, as in graph-based disassembly and type inference
\cite{DeepDi2022,Zhu2024}.
Other low-frequency families also reflect narrow task formulations:
prefix-tree learning appears in function-start recognition, while a
learned tree transducer appears in assembly-to-IR lifting
\cite{Bao2014,Hasabnis2016}.
Architecture family therefore provides an initial signal of task form
rather than a standalone explanation of technical design: later
source-like stages are increasingly Transformer-centered, while upstream
and structure-heavy tasks still often use architectures that operate
over local, sequential, graph, or symbolic evidence.

\subsubsection*{Model-use strategies}
We next distinguish how a model is fitted to a decompilation task:
training a task-specific model, adapting a reusable backbone, or using a
fixed model through prompting.
We use \emph{model-use strategy} to cover all three cases: training from
scratch, adapting a reusable or pre-trained backbone, and using a fixed
model through prompting.
This model-use view is closer than architecture family to the question
of how ML is made useful for decompilation.
In the 116-study ML comparison subset, each study has one primary
model-use strategy: 62 are coded as from-scratch training, 38 as model
adaptation, and 16 as prompting-based use, split across
analysis-augmented, feedback-guided, and context-augmented prompting.
This distribution does not show a simple chronological replacement of
custom training by prompting.
From-scratch training remains the largest category, especially in custom
supervised settings where the authors construct task-specific labels or
paired training data and train a dedicated model for boundaries, types,
names, or whole-function translation.
Examples include neural function recognition, graph-based disassembly,
and a small Transformer decompiler trained for assembly-to-C generation
\cite{Shin2015,DeepDi2022,SLaDe2024}.
This coding rule matters because from-scratch training also includes
boundary cases in which a model is pre-trained and fine-tuned within the
same study: the extraction schema reserves ``model adaptation'' for reuse
or adaptation of an external or reusable backbone.

Model adaptation and prompting become more visible where the target
artifact can be represented as lexical information, pseudocode, or
source-like code for code-oriented language models.
The two modes differ in how reuse is performed: model adaptation changes
the model through continued pre-training or fine-tuning, whereas
prompting keeps the model fixed and changes the information supplied at
inference time.
Model adaptation includes continued pre-training or fine-tuning of
reusable backbones for binary-specific tasks, such as pseudocode-centric
binary information inference, large LLM decompilation, and
domain-adapted function-name generation
\cite{HexT52023,LLM4Decompile2024,Jiang2025}.
Prompting, by contrast, appears mainly as a wrapper around existing
LLMs rather than as a trained decompiler.
In P8, prominent prompting examples are not merely direct prompts: they add
retrieved examples, static-analysis structure, compiler diagnostics, or
runtime feedback to steer refinement and repair
\cite{DecLLM2025,PseudoFix2025,HELIOS2026}.
Model-use strategy therefore qualifies the architectural pattern
described above.
Transformer-based systems in this corpus are not uniform: some are
trained as task-specific models, some adapt reusable code models, and
some rely on prompt-time analysis or feedback.
The useful distinction is thus not a performance ordering among these
modes, but how each mode connects available supervision, recovered
artifacts, and decompilation-specific constraints.

\subsubsection*{Reusable model backbones}
Table~\ref{tab:ml-model-names} reports canonical reusable or external
model names in the ML comparison subset.
Such names are recorded for 61 of the 116 studies, with 95 model-name
occurrences across 39 distinct names.
The most frequent names include GPT-4, CodeLlama, LLM4Decompile,
GPT-3.5, Qwen2.5-Coder, and CodeT5.
This pattern indicates substantial reuse of general-purpose LLMs and
code-oriented foundation models, alongside custom task-specific models.
However, model names are less informative than model-use strategies by
themselves.
The same backbone can be used through prompting, continued pre-training,
fine-tuning, retrieval support, or as a baseline, and different papers
report model names at different levels of specificity.
Reusable backbones are therefore best read as evidence of model reuse,
not as a standalone technical category: they indicate which external
model ecosystems the field draws on, while the model-use analysis
explains how those models are made task-specific.

\input{tables/table_ml_model_names}

\subsubsection*{Execution platforms}
Here, execution platform refers to the target binary or bytecode
environment rather than the hardware used to train or run the model.
Table~\ref{tab:ml-platform-profile} shows that the ML comparison subset
is strongly native-binary centered: 97 of the 116 studies target native
binaries.
The remaining studies are distributed across smaller platform-specific
settings, including EVM, WebAssembly, JVM, and an Other group covering
CUDA/PTX, Move VM, Pharo VM, and Python VM targets.
This distribution does not by itself determine the technical design of a
method, but it is an important background condition for interpreting
design and evaluation claims.
Studies targeting native binaries may face different artifacts,
toolchains, and correctness oracles from those targeting smart-contract
bytecode, WebAssembly, CUDA/PTX, or high-level VM bytecode.
Execution platform is therefore best read as a contextual design axis:
it shows where ML-based decompilation evidence is concentrated and where
the evidence remains platform-specific.

\input{tables/table_ml_platform_profile}

\paragraph*{Answer to RQ2.1.}
ML-based studies adapt ML to decompilation through combinations of task
form, model-use strategy, reusable model ecosystem, and target execution
platform rather than through a single dominant technical pattern.
Transformer-family models are most visible where the task operates over
lexical, pseudocode, or source-like artifacts, while upstream and
structure-heavy tasks also use sequence, graph, or specialized models
that operate over local program evidence.
The central distinction is how these models are made task-specific:
many studies train dedicated models from task-specific labels or
paired data, whereas others adapt reusable code-oriented backbones or
keep the model fixed and supply decompilation-specific context at
prompt time.
Reusable model names show substantial reuse of LLM and code-model
ecosystems, but those names only become informative when read together
with model-use strategy and task formulation.
Finally, the evidence base remains strongly native-binary centered, with
smaller platform-specific lines for smart-contract bytecode,
WebAssembly, CUDA/PTX, and high-level VM bytecode.
Thus, RQ2.1 is best answered by viewing technical design as task- and
artifact-conditioned model use, not as a shift toward one ML
architecture or prompting style.

\FloatBarrier

\subsection{Evaluation Evidence and Metrics (RQ2.2)}
\label{sec:focused-review-evaluation-metrics}

Evaluating ML-based decompilation is not a matter of constructing a
single leaderboard.
The studies in the comparison subset report many metric names, but those
metrics do not all measure the same kind of improvement.
For example, matching a recovered function boundary, predicting a type
or name, producing text close to a reference output, recompiling
generated code, passing execution tests, and supporting a downstream
analysis all provide different forms of evidence.
We therefore group reported metrics by the kind of evaluation evidence
they provide for decompilation outputs.
This normalization lets us ask what each metric is being used to support
rather than treating metric names as directly comparable scores.

\begin{figure}[pos=t]
  \centering
  \includegraphics[width=0.98\columnwidth]{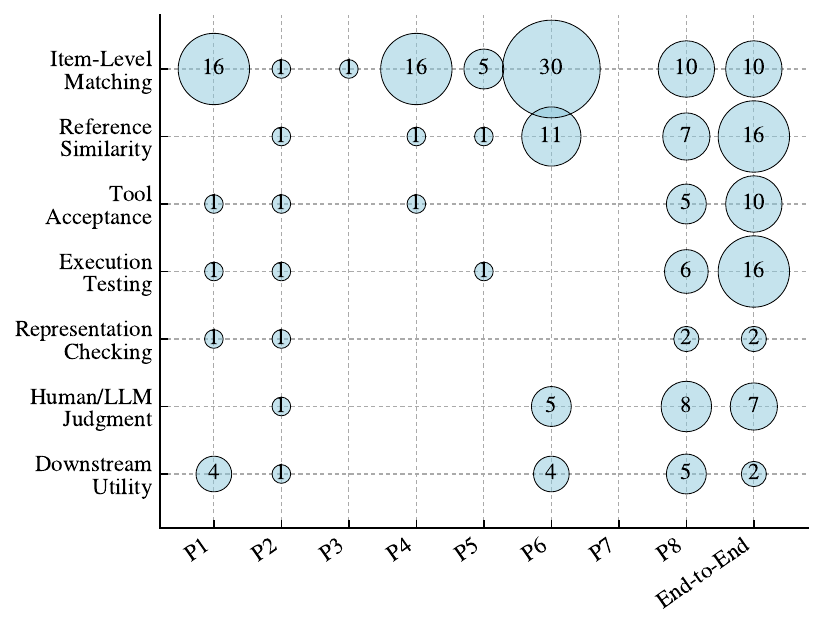}
  \caption{Distribution of evaluation-evidence types across pipeline
  stages. Counts are paper-level counts within the
  116-study comparison subset.}
  \label{fig:ml-evaluation-evidence-heatmap}
  \vspace{-0.75\baselineskip}
\end{figure}

The 116-study comparison subset contains 450 reported metric entries,
with at least one metric entry captured for each study.
Because a single study may report several metrics,
Table~\ref{tab:metric-examples-by-type} and
Figure~\ref{fig:ml-evaluation-evidence-heatmap} emphasize paper-level
evidence-type counts when summarizing the distribution, while
metric-entry counts are used only to describe the range of reported
metric names.
Table~\ref{tab:metric-examples-by-type} lists frequent normalized
metrics within each evidence type, and
Figure~\ref{fig:ml-evaluation-evidence-heatmap} shows where those
evidence types appear across pipeline stages.
These evidence types are not ordered as a scale of evidential strength,
but they form a rough progression from local item agreement, through
textual and tool-acceptance properties of generated outputs, toward
behavioral, representational, human-facing, and downstream-use
evidence.
The rest of this subsection reads each evidence type in the same way:
what it measures, which decompilation task forms it fits, and what kind
of improvement claim it cannot establish by itself.

\input{tables/table_metric_examples_by_type}

\noindent\textbf{Item-level matching.}
Item-level matching measures whether a recovered local artifact matches
a reference item, such as a function boundary, instruction label, type,
identifier, or classification target.
It is the most common evidence type in the comparison subset, appearing
in 89 papers.
At the pipeline level, Figure~\ref{fig:ml-evaluation-evidence-heatmap}
shows that item-level matching appears across the pipeline, with the
largest paper-level counts in P6, P1, and P4.
It appears in 30 P6 studies, 16 P1 studies, 16 P4 studies, 10 P8
studies, 10 End-to-End studies, five P5 studies, and one study each in
P2 and P3.
The frequent metrics in this group are local matching and prediction
metrics: precision, recall, F1-score, accuracy, and smaller numbers of
top-$k$ or exact-match variants.
This evidence type is most natural when a decompilation task can be
decomposed into individually checkable decisions rather than requiring
evaluation of a whole generated program.

The same evidence form appears across several different task settings.
Function-boundary recovery can be evaluated by matching predicted
function entries or boundaries against references; type recovery can be
evaluated by matching recovered type labels against debug or source-level
references; and symbol or name recovery can be evaluated by matching
predicted tokens or labels against original names
\cite{Shin2015,StateFormer2021,Debin2018}.
This makes item-level matching a direct source of evidence for localized
prediction quality, and this pattern is consistent with its frequency
across upstream recovery, type recovery, and name recovery.
At the same time, such metrics do not by themselves establish semantic
equivalence, readability, recompilability, or executable behavior of a
complete decompiler output.

\noindent\textbf{Reference similarity.}
Reference similarity measures textual, structural, or semantic proximity
between a generated or refined artifact and an available reference or
reported source-like property.
It appears in 37 papers and is concentrated in output-facing settings:
16 End-to-End studies, 11 P6 studies, and seven P8 studies use this
evidence type, with one study each in P2, P4, and P5.
The frequent metrics include BLEU, ROUGE, character error rate, edit
distance, edit similarity, CodeBLEU, and embedding- or
semantics-oriented similarity measures.
This evidence type fits tasks where the evaluated artifact has a
textual, structural, or source-like form rather than a single local
label.

Its use differs across task settings.
In P6, identifier-recovery work can compare predicted names or lexical
artifacts against source-level references using character-level or
token-similarity measures~\cite{DIRECT2021}.
In P8, refinement work can compare pseudocode against textual,
structural, or source-like properties of a reference
output~\cite{PseudoFix2025}.
In End-to-End generation, edit-distance and code-similarity measures can
compare generated code with reference source-like output
\cite{SLaDe2024,LLM4Decompile2024}.
This makes reference similarity useful evidence that an output has moved
toward an available reference or reported source-like property, but it
does not by itself establish semantic equivalence, recompilability, or
executable behavior, especially because multiple source-level
renderings may be valid for the same binary behavior.

\noindent\textbf{Tool acceptance.}
Tool acceptance measures whether a generated, refined, or
analysis-facing artifact or tool interaction is accepted by an external
toolchain or analysis environment.
It is not a behavioral check on the recovered program itself; rather, it
asks whether the output satisfies the syntactic, interface, or tool-use
constraints required by a compiler, recompiler, linker, decompiler
service, or downstream analysis tool.
It appears in 18 papers and is concentrated in output-facing settings:
10 End-to-End studies and five P8 studies use this evidence type, with
one study each in P1, P2, and P4.
The reported metrics include recompilation success, compilation success,
syntactic completeness, syntax-error rates, tool-applicable output
rates, executable linkability, and tool-call correctness.

The role of this evidence type varies with the evaluated artifact.
In output-refinement and End-to-End settings, studies may report whether
generated code can be recompiled, linked, or accepted by a compiler
before separately testing its runtime behavior
\cite{PCodeTrans2026,HELIOS2026}.
In other settings, tool acceptance can capture whether an analysis
output clears syntax-level or tool-level checks before stricter
equivalence or downstream analyses are considered~\cite{PyLingual2025}.
This makes tool acceptance useful evidence for syntactic and
toolchain-level usability, but it does not by itself establish semantic
equivalence or executable behavior.

\noindent\textbf{Execution-based testing.}
Execution-based testing measures whether a generated, refined, or
recovered artifact behaves as expected when it is executed or checked
against execution-derived observations.
It appears in 25 papers and is concentrated in output-facing settings:
16 End-to-End studies and six P8 studies use this evidence type, with
one study each in P1, P2, and P5.
The reported metrics include re-execution success, output consistency,
pass@k-style measures, runtime correctness, test-case pass rates,
dynamic verification, exploit reproduction, and execution-path accuracy.
Unlike tool acceptance, which asks whether an artifact is accepted by a
toolchain, execution-based testing asks whether the accepted or
generated artifact exhibits expected behavior under some execution
condition.

This evidence type is especially natural when the decompilation output
can be run against inputs, linked into an executable setting, or checked
against expected runtime observations.
End-to-End systems may report output consistency, IO accuracy, or
re-executability on compiled benchmark functions
\cite{SLaDe2024,LLM4Decompile2024}, while output-refinement systems may
first obtain compilable code and then test whether the compiled or
recompiled artifact matches expected runtime behavior
\cite{PCodeTrans2026,HELIOS2026}.
Execution-based testing is therefore closer to whole-output behavior
than reference similarity or tool acceptance, but it still depends on
test coverage, input harnesses, and target execution environments, and
does not by itself establish full semantic equivalence.

\noindent\textbf{Representation-based checking.}
Representation-based checking measures whether a recovered or generated
artifact satisfies structural, semantic, or equivalence constraints in a
specified representation or checking procedure.
It is rare in the comparison subset, appearing in six papers, with
metrics such as equivalence-check pass rate, soundness, loop-back test
pass rate, semantic error rate, D-SCORE, and structural-constraint
violation rate.
The category appears where the evaluation can rely on an explicit
representation-level oracle: learned lifting can test soundness and
round-trip properties of lifted instructions~\cite{Hasabnis2016};
neural disassembly can report violations of global structural
constraints~\cite{Tady2025}; and End-to-End decompilation can use
explicitly defined perfect-decompilation or equivalence-style checks for
generated artifacts~\cite{PyLingual2025}.
This evidence type is closer to structural or semantic validity than
plain textual similarity or tool acceptance, but it depends on the
availability and scope of the representation, checker, or oracle, and
does not by itself establish general semantic equivalence across all
inputs and program contexts.

\noindent\textbf{Human/LLM judgment.}
Human/LLM judgment measures qualities of decompilation outputs that are
hard to reduce to exact labels, textual similarity, or execution tests,
such as readability, semantic plausibility, contextual relevance,
comment usefulness, and user preference.
It appears in 21 papers and is concentrated in human-facing stages:
eight P8 studies, seven End-to-End studies, five P6 studies, and one P2
study use this evidence type.
The frequent metrics include code semantic correctness rate, user-study
preference, readability score, context-relevance score, and user-study
score.
These metrics are natural when the claimed improvement concerns how a
reverse engineer or an LLM judge interprets a source-like artifact rather
than whether a single item matches a reference.
Examples include human ratings of readability, consistency, and
reasonability for pseudocode refactoring~\cite{PseudoFix2025},
readability scoring for LLM-based decompilation~\cite{LLM4Decompile2024},
and user-study or comment-quality measures for decompiler-output
optimization~\cite{DeGPT2024}.
This evidence type is closer to human-facing or judged output quality
than reference similarity alone, but it depends on the evaluator, rubric,
prompt or judge model, and user-study design, and does not by itself
establish semantic equivalence or executable behavior.

\noindent\textbf{Downstream utility.}
Downstream utility measures whether a recovered or generated artifact
improves a later analysis, audit, security, or user task.
It appears in 16 papers, with a different distribution from
human/LLM judgment: five P8 studies, four P1 studies, four P6 studies,
two End-to-End studies, and one P2 study use this evidence type.
The frequent metrics include accuracy, AICT, precision, recall, task
completion time, and F1-score, but the category is better understood by
the task setting than by metric names alone.
In upstream recovery, improved call-graph information can be evaluated by
its effect on call-graph-dependent analyses such as binary similarity
and fuzzing~\cite{Callee2023}.
In output-refinement settings, recompilable decompiler output can be
evaluated by whether it enables CodeQL-based vulnerability analysis
\cite{DecLLM2025}.
In smart-contract decompilation, generated source-like artifacts can be
connected to attack reproduction tasks~\cite{DiSCo2025}.
This evidence type is among the most directly tied to practical use in
this subsection, but it depends strongly on the downstream task, tool,
benchmark, and user setting, and does not by itself establish general
output quality or semantic equivalence.

\noindent\textbf{Dataset and benchmark context.}
The interpretation of evaluation evidence also depends on the data on
which metrics are applied.
We do not treat dataset or benchmark names as a complete corpus-wide
quantitative axis, because the same name can refer to different versions, splits,
compiler settings, architectures, optimization levels, stripping or
obfuscation choices, and execution harnesses.
Nevertheless, the extracted records include named datasets and
benchmarks that are useful as qualitative context for interpreting the
evaluation landscape.
Table~\ref{tab:dataset-benchmark-names} lists illustrative examples.
The table is not a ranking of dataset use, nor evidence that studies
followed identical benchmark protocols.
The examples suggest that ML-based decompilation studies draw from
several different evaluation-data traditions: reusable binary or
decompilation benchmarks, programming-task benchmarks adapted into
compiled settings, decompiler-output corpora for local recovery, and
domain-specific corpora such as EVM smart contracts.
This reinforces the point that evaluation heterogeneity comes not only
from metric choice, but also from the construction and role of the data
to which those metrics are applied.

\begin{table}[pos=t]
\centering
\caption{Illustrative dataset and benchmark resources mentioned in
ML-based decompilation studies. The table is qualitative rather than
quantitative; entries are not ranked by frequency.}
\label{tab:dataset-benchmark-names}
\begin{tabularx}{\columnwidth}{@{}>{\raggedright\arraybackslash}p{0.48\columnwidth}>{\raggedright\arraybackslash}X@{}}
\toprule
Benchmark/Dataset & Languages \\
\midrule
ALLSTAR & Mixed \\
\addlinespace[0.25ex]
Decompile-Eval & C/C++ \\
\addlinespace[0.25ex]
DIRE dataset & C \\
\addlinespace[0.25ex]
DIRT dataset & C \\
\addlinespace[0.25ex]
ExeBench & C \\
\addlinespace[0.25ex]
GNU Coreutils & C \\
\addlinespace[0.25ex]
HumanEval & Python \\
\addlinespace[0.25ex]
HumanEval-Decompile & C \\
\addlinespace[0.25ex]
MBPP & Python, C/C++ \\
\addlinespace[0.25ex]
Nero & C \\
\addlinespace[0.25ex]
neural-FEBI & Solidity \\
\addlinespace[0.25ex]
TYDA & C/C++ \\
\bottomrule
\end{tabularx}
\end{table}

\paragraph*{Answer to RQ2.2.}
ML-based decompilation studies most often support improvement claims
through item-level matching.
In many studies, improvement is shown by whether a local
recovery target, such as a function boundary, type, identifier, or
classification label, matches a reference.
This means that the dominant evaluation evidence in the comparison
subset is localized reference agreement rather than whole-output
semantic equivalence.

Output-facing studies use a broader evidence profile.
In P8 and End-to-End studies, where the recovered artifact is closer to
pseudocode, source code, or an executable program, reference similarity,
tool acceptance, and execution-based testing become more visible.
These evidence types support claims that generated or repaired outputs
are closer to references, accepted by toolchains, or consistent with
expected runtime behavior.

Other evidence types address semantic, human-facing, or practical claims
more directly.
Representation-based checking provides explicit structural or
equivalence-style evidence, while human/LLM judgment and downstream
utility capture readability, judged quality, analyst-facing usefulness,
or effects on later analysis tasks.
Across these evidence types, dataset and benchmark context is an
additional source of heterogeneity.
Some named resources recur, but many studies adapt programming
benchmarks, rebuild common source packages, or construct
platform-specific and transformed corpora under different experimental
conditions.
Thus, metric names alone do not determine comparability: the same
accuracy, similarity, or execution-based metric can support different
claims depending on the evaluation data and oracle used.
Overall, RQ2.2 shows that evaluation in ML-based decompilation should be
read as task-specific evidence profiles rather than as a single
leaderboard-style comparison across all studies.

\subsection{Summary of ML-Focused Findings for RQ2}
\label{sec:focused-review-synthesis}

In answer to RQ2, ML-based decompilation studies are best
characterized by coupled technical-design and evaluation-evidence
profiles.
They do not form a single technical trajectory in which one model family
or prompting style replaces the classical pipeline.
Instead, both design choices and evaluation evidence vary with the
artifact being recovered.
Local recovery tasks are commonly framed around task-specific labels and
are often evaluated through item-level reference agreement.
Lexical, pseudocode, and source-like tasks more often draw on reusable
or adapted language-model backbones, and their claims are supported by
more heterogeneous evidence, including reference similarity, tool
acceptance, execution behavior, human/LLM judgment, and downstream
utility.
The evidence base is also platform-conditioned: it remains strongly
native-binary centered, while EVM, WebAssembly, CUDA/PTX, and other VM
targets appear as smaller platform-specific lines.
The common pattern is therefore a task-, artifact-, and
platform-conditioned design and evidence profile.
ML is made useful by fitting a model, prompt, or reusable backbone to a
particular decompilation artifact, and improvement claims are supported
by evaluation evidence suited to that artifact rather than by a common
cross-study score.

\noindent\textbf{Bridge to discussion.}
This synthesis sets up the discussion by explaining why ML-based
decompilation should not be treated as a single ranked family.
The focused review shows where ML extends the decompilation pipeline
toward source-level completion, naming, refinement, and whole-function
generation, but it also shows that evidence for these claims remains
uneven across task forms.
Local prediction tasks are often supported by direct reference
agreement, whereas broader claims about semantic fidelity,
recompilability, analyst usefulness, or downstream value require
different and less uniformly available evidence.
Section~\ref{sec:discussion} therefore turns from classification to
interpretation: how ML complements classical analysis, where output
readability and semantic faithfulness diverge, and what kinds of
evaluation practice are needed for stronger cross-study comparisons.

%% file: tables/table_ml_model_names.tex
\begin{table}[pos=t]
\centering
\small
\caption{Most frequent canonical reusable model names in the 116-study ML
comparison subset (canonical names mentioned in at least three studies).
Counts are paper-level model mentions after canonicalization of reusable
or external model names only.}
\label{tab:ml-model-names}
\begin{tabular}{lr}
\toprule
Canonical model & Studies \\
\midrule
GPT-4 & 8 \\
CodeLlama & 7 \\
LLM4Decompile & 7 \\
GPT-3.5 & 6 \\
Qwen2.5-Coder & 5 \\
CodeT5 & 4 \\
DeepSeek-V3 & 4 \\
Llama-3 & 4 \\
BERT & 3 \\
DeepSeek-Coder & 3 \\
Gemini & 3 \\
GPT-4o & 3 \\
\bottomrule
\end{tabular}
\end{table}

%% file: tables/table_ml_platform_profile.tex
\begin{table}[pos=t]
\centering
\small
\caption{Execution-platform profile of the harmonized comparison subset
(116 studies).}
\label{tab:ml-platform-profile}
\begin{tabularx}{\columnwidth}{@{}>{\raggedright\arraybackslash}Xr@{}}
\toprule
Platform & Studies \\
\midrule
Native & 97 \\
EVM & 8 \\
WebAssembly & 4 \\
JVM & 3 \\
Other (CUDA/PTX, Move VM, Pharo VM, Python VM) & 4 \\
\bottomrule
\end{tabularx}
\end{table}

%% file: tables/table_metric_examples_by_type.tex
\begin{table*}[pos=t]
\centering
\caption{Normalized metrics grouped by evaluation evidence type. The table
shows metrics used by at least two papers. The Papers column gives the number
of papers using each evidence type; counts in parentheses are paper-level counts
for individual normalized metrics. One-paper metric names are retained in the
extraction data but omitted from this table for readability.}
\label{tab:metric-examples-by-type}
\begin{tabularx}{0.96\textwidth}{@{}p{0.27\textwidth}r>{\raggedright\arraybackslash}X@{}}
\toprule
Evidence type & Papers & Frequent normalized metrics \\
\midrule
Item-Level Matching & 89 & Precision (55), Recall (54), F1-score (52), Accuracy (33), Accuracy@k (8), Exact Match (5), Recall@k (4), AUROC (3), Function recovery completeness (2), Precision@k (2) \\
\addlinespace[0.25ex]
Reference Similarity & 37 & ROUGE (7), BLEU (6), Character Error Rate (6), Edit distance (6), Edit similarity (6), Accuracy (3), CodeBLEU (3), METEOR (3), BERTScore (2), chrF (2), Cyclomatic complexity (2), Embedding similarity (2), Semantic similarity (2), Type Prefix Score (2), Variable semantic similarity score (2) \\
\addlinespace[0.25ex]
Tool Acceptance & 18 & Recompilation success rate (10), Compilation success rate (4) \\
\addlinespace[0.25ex]
Execution-Based Testing & 25 & Re-execution success rate (17), Output consistency rate (4), Pass@k (2) \\
\addlinespace[0.25ex]
Representation-Based Checking & 6 & Equivalence check pass rate (4) \\
\addlinespace[0.25ex]
Human/LLM Judgment & 21 & Code semantic correctness rate (4), User study preference (4), Readability score (3), Context relevance score (2), User study score (2) \\
\addlinespace[0.25ex]
Downstream Utility & 16 & Accuracy (5), AICT (Average indirect call targets) (3), Precision (3), Recall (3), Task completion time (3), F1-score (2) \\
\bottomrule
\end{tabularx}
\end{table*}

%% file: sections/7_discussion.tex
\section{Discussion}\label{sec:discussion}

This section does not introduce a separate research question.
Instead, it interprets the answers to \textbf{RQ1} and
\textbf{RQ2} together, drawing out the cross-cutting implications of
the landscape in Section~\ref{sec:landscape} and the focused review in
Section~\ref{sec:focused-review}.
The mapping in Section~\ref{sec:landscape} and the comparative review
in Section~\ref{sec:focused-review} together reveal a field with uneven
development across its pipeline and uneven evaluation practices.
Section~\ref{sec:focused-review-synthesis} summarized this unevenness in
terms of a task-, artifact-, and platform-conditioned relationship
between technical design and evaluation evidence.
This section interprets that relationship: why ML enters some parts of
the pipeline more readily than others, why hybrid designs remain
important, why cross-study comparison is difficult, and what kinds of
future work follow from the evidence profile.

\subsection{Structural vs.\ Semantic Recovery}

A useful way to read the current field is through recoverability, not
merely through model family.
Stages such as front-end recovery, lifting, and control-flow structuring
reconstruct artifacts whose relation to the input binary remains
comparatively constrained.
For these tasks, compiler-theoretic reasoning, validity checks, and
static invariants remain powerful because the search space is structured
enough that preserving exact low-level meaning is often more important
than generating plausible source-level forms.

Tasks such as type recovery, name recovery, output refinement, and
End-to-End generation differ because they target information that
compilation has partly or wholly discarded.
Variable names, richer type semantics, readability-oriented repair, and
whole-function source reconstruction all require some form of semantic
completion under irreversible information loss.
That does not make them unconstrained, but it does make them less
amenable to exact inversion and more dependent on learned priors about
source code, naming conventions, and plausible program behavior.

This helps explain why the map in Section~\ref{sec:landscape} and the
focused review in Section~\ref{sec:focused-review} show different
method profiles across artifacts.
ML becomes most visible where the target is not a direct structural
property of the binary, but a source-level or human-facing artifact that
must be inferred, ranked, rewritten, or generated.
This suggests that the resulting structure--semantics divide is not only
a matter of current model capacity.
It also reflects a deeper difference between recovering constrained
program structure and completing information that compilation has made
ambiguous or unavailable.

\subsection{The Case for Hybrid Designs}

The comparative review does not support a simple replacement
narrative; it instead suggests durable complementarity.
Across the ML-focused subset, learned components are commonly described
in relation to artifacts produced by disassemblers, decompilers, static
analyses, or intermediate representations.
They then apply prediction, adaptation, or prompting where semantic
completion, naming, rewriting, or source-like rendering becomes
valuable.
In this sense, the field appears less like a transition toward a single
monolithic ML decompiler than like repeated divisions of labor between
constrained analysis and flexible inference.

This framing makes hybrid architectures appear less like a temporary
engineering compromise and more like a recurring practical pattern in
the current evidence.
If some components are better suited to recovering constrained
structure, and others to ranking or generating plausible source-level
completions, then the design problem shifts from choosing one paradigm
over the other to designing interfaces between them.
Those interfaces need to carry structural constraints forward, expose
uncertainty where it matters, and leave room for validation or
re-execution downstream.
A staged pipeline in which components can be inspected or verified
independently is therefore consistent with the task-conditioned pattern
observed in the current evidence.

\subsection{Why Cross-Study Comparison Remains Difficult}

Evaluation remains fragmented not only because the ML-based literature
is recent, but also because decompilation targets heterogeneous
artifacts while lacking widely shared faithfulness oracles.
Types, names, structured code, and executable programs are not merely
different outputs; they embody different notions of success, and the
evaluation evidence used for them overlaps only partially.
Within a stage, corpus construction, optimization settings, toolchain
choices, and label-space design further affect what a reported score
actually means.

The deeper issue is that readability, re-executability, exact match,
and semantic faithfulness are related but non-identical targets.
A readable identifier can still miss the intended semantics; an
executable function can still diverge from the original program; and a
strict verifier can reject outputs that are semantically acceptable but
not identical in form.
This helps explain why the evidence types in
Section~\ref{sec:focused-review-evaluation-metrics} are claim-specific,
not interchangeable.
Item-level matching can support claims about local recovery quality,
reference similarity can support claims about proximity to a reported
source-like artifact, and execution-based testing can support claims
about behavior under particular tests or harnesses.
None of these automatically supplies a common scale for the full
decompilation problem.

Shared benchmarks and reporting standards still matter, but they will
not by themselves resolve the comparison problem.
A recurring need is broader access to scalable semantic checks and a
clearer separation between evidence for local agreement, legibility,
tool acceptance, executable behavior, human-facing usefulness, and
faithful recovery.
Until those distinctions are made explicit, cross-paper comparisons will
remain informative but only partly commensurate.

\subsection{Implications for Research and Practice}

For practitioners, the mapping and focused review carry a direct
message: ML-based components are currently strongest as additions to,
not replacements for, a classically grounded decompilation workflow.
Recovered names, types, and refined outputs should be treated as
plausible hypotheses whose reliability depends on the evidence type
behind them; item-level accuracy on a benchmark does not guarantee
semantic faithfulness on an analyst's binary, and readability gains do
not imply behavioral equivalence.
Tool builders can use the pipeline map to decide where learned
components add value and where classical invariants should remain
authoritative.

For researchers, the most useful roadmap is the one suggested by the
current evidence, not the one implied by general enthusiasm for larger
models.
Four directions stand out.
First, scalable semantic verification or equivalence testing would
complement the local and textual evidence that currently dominates many
ML-based evaluations.
Such checks would make it easier to evaluate faithful recovery directly
rather than inferring it from reference similarity, readability, or proxy
executability.
Second, robustness and generalization should be tested more systematically
across compilers, optimization levels, architectures, stripped settings,
and obfuscation, because the focused review shows that evidence remains
strongly conditioned by platform and evaluation settings, with the
current evidence base still centered mainly on native binaries.

Third, decompilation quality should be evaluated more often through the
downstream tasks it is meant to support, including vulnerability
analysis, malware investigation, and other reverse-engineering
workflows.
That would complement stage-local scores with evidence about practical
utility.
Fourth, hybrid interface design should receive more direct attention:
if classical and learned components remain complementary, then the
exchange of structural facts, uncertainty, retrieval context, execution
feedback, and verification signals becomes central rather than
auxiliary.
This agenda keeps the discussion tied to the observed evidence profile
rather than expanding into a broad catalog of possible gaps.

%% file: sections/8_threats.tex
\section{Threats to Validity}\label{sec:threats}

\paragraph*{Search completeness.}
Our corpus depends on the coverage of the search strings, databases, and
citation-tracing procedure.
Although the query set was designed to combine direct decompilation
terms with complementary binary-context terms, studies that describe
decompilation-adjacent problems using terminology outside these query
families may have been missed.
Recent preprints may also be missing if they were not indexed or
retrieved before the search cutoff.
We mitigated this risk by searching seven databases, including arXiv,
performing one round of forward and backward citation tracing, and
checking that known relevant studies were captured.
Citation tracing was limited to a single round for tractability at the
scale of this corpus; additional rounds could surface further studies,
particularly in older or less consistently indexed literature.
Although this procedure reduces the risk of missing studies that use
different terminology or are less consistently indexed, the corpus should
not be read as a saturated bibliography of all adjacent binary-analysis
work.

\paragraph*{Screening and extraction reliability.}
The initial screening phase involved two reviewers: relevance judgments
on candidate records were independently checked by a second reviewer,
with disagreements resolved by discussion.
However, inter-rater agreement was not quantified (e.g., with Cohen's
kappa), so the reliability of the screening decisions cannot be
reported statistically.
Classification and data extraction beyond this screening check were
performed by a single reviewer.
This introduces a risk of subjective judgment in borderline
inclusion/exclusion decisions, publication-type decisions, stage
assignments, and method tags.
To make these decisions auditable, we retain the complete per-study
classification and extraction records in the replication package.
In the final mapping, all primary studies have resolved stage
assignments; two secondary records retain provisional stage annotations,
but they do not affect the primary-stage counts reported in the paper.
The results should therefore be read with this predominantly
single-reviewer design in mind, even though the released records support
independent inspection.

\paragraph*{Taxonomy validity.}
The P1--P8 pipeline, together with the complementary End-to-End and
Evaluation categories, is an analytical reference model rather than a
claim that all studies fit a strict implementation sequence.
It is derived from Cifuentes's classical framework and updated with
subsequent work, but alternative decompositions are possible.
Some studies straddle stage boundaries or combine classical and
ML-based techniques.
For corpus-level mapping, each primary study is assigned one primary
taxonomy category and one primary method tag, with secondary roles
retained in coding notes for interpretation.
This makes the heatmap readable and supports comparison across the
corpus, but it also simplifies boundary cases.
The stage profiles and heatmap should therefore be interpreted as a
mapping of dominant contribution positions, not as a complete account of
every role played by each study.

\paragraph*{ML-focused extraction validity.}
The focused review depends on additional extraction and normalization for
the 116-study ML comparison subset.
Architecture families, model-use strategies, reusable model names, execution
platforms, metrics, and evaluation-evidence types are not always
reported at the same level of detail across papers.
Reusable model names may be omitted or reported at different levels of
specificity, and model-use strategies can be ambiguous when a study
combines pre-training, fine-tuning, prompting, and analysis-derived
context.
The metric normalization in Section~\ref{sec:focused-review-evaluation-metrics}
also requires interpretation: the same metric name can support different
claims depending on whether it is used for a local label, a source-like
artifact, a tool-accepted output, or an executed program.
We mitigated this risk by making the grouping rules explicit, preserving
the extracted records, and reporting paper-level evidence-type counts
separately from metric-entry counts.
Alternative normalizations could still shift emphasis among design and
evaluation categories.

\paragraph*{Temporal validity.}
The corpus covers publications through early 2026 and should be read as
a snapshot of the field at that cutoff.
This is especially important for ML-trained and prompted LLM-based
decompilation, where new models, preprints, and tool integrations appear
rapidly.
Studies published or revised after the cutoff are not included, and the
relative prominence of prompted, adapted, or platform-specific work may
change as the literature evolves.

\paragraph*{Publication and reporting bias.}
Systematic reviews inherit the publication bias of their source
literature: negative results, failed approaches, and inconclusive
comparisons are less likely to be published.
The corpus also includes 27~preprints that have not undergone peer
review; excluding them would understate the fast-moving ML-based
literature, but their findings have not been independently vetted and
should be weighted accordingly.
Our synthesis also depends on what papers report.
Model details, prompts, datasets, evaluation settings, and unsuccessful
evaluation attempts may be absent or described only partially, which
limits what can be extracted and compared.
The improvements reported in the ML-based studies reviewed in
Section~\ref{sec:focused-review} should therefore be read as reported
claims under the evaluation settings used by those studies, not as a
direct estimate of typical performance in practice.

%% file: sections/9_conclusion.tex
\section{Conclusion}\label{sec:conclusion}

This paper presented a systematic survey of 405 decompilation studies
spanning 1973--2026, including 394 primary studies analyzed through a
field-level mapping (RQ1) and a focused comparative review of ML-based
approaches (RQ2).

The mapping shows that decompilation research is unevenly distributed
over time and across the pipeline-centered taxonomy.
Front-end recovery, IR lifting, and control-flow structuring remain
dominated by classical techniques grounded in compiler theory and
formal methods.
Type recovery is the clearest mixed region, while data-structure
recovery remains more classical-leaning but includes targeted ML and
hybrid entries.
Symbol/name recovery and output refinement are the pipeline stages where
ML-based work is most concentrated, while End-to-End translation forms a
complementary cross-stage category with a similarly strong ML presence.
The resulting map provides positional evidence: classical and ML-based
approaches are not competing on a single common scale, but are located
differently across the decompilation pipeline.

The ML-focused review further indicates that ML-based decompilation is
likewise not a single technical trajectory.
Technical design varies with the task, the recovered artifact, the
model-use strategy, the reusable model ecosystem, and the target execution
platform.
Transformer and LLM-based approaches are most visible where the task
operates over lexical, pseudocode, or source-like artifacts, while other
sequence, graph, and specialized models remain relevant for more local or
structure-heavy tasks.
Evaluation evidence is similarly task-specific.
Many studies support improvement claims through item-level matching,
whereas output-facing and End-to-End studies draw on a broader evidence
profile that includes reference similarity, tool acceptance,
execution-based testing, human/LLM judgment, and downstream utility.
Taken together, these findings indicate that ML-based decompilation
should be interpreted by considering its pipeline position, technical
design, and evaluation evidence together, rather than as a single
leaderboard of methods.

These findings point toward four directions for future work.
First, scalable semantic verification and equivalence testing would
complement the local and textual evidence that currently dominates many
ML-based evaluations.
Second, robustness testing should be broadened across compilers,
optimization levels, architectures, stripped settings, obfuscation, and
execution platforms.
Third, evaluating decompilation in the context of downstream
reverse-engineering tasks would connect intermediate output quality to
practical utility.
Fourth, hybrid interface design should receive more direct attention
given the observed complementarity between classical analyses and
learned components in building trustworthy decompilation pipelines.